\documentclass[aps,prb,reprint,floatfix,superscriptaddress]{revtex4-2}

\usepackage[english]{babel} % Language hyphenation and typographical rules
\usepackage{hyperref}
\usepackage{amssymb,amsmath}
\usepackage{graphicx}
\usepackage{float}
\usepackage{url}
\usepackage{multirow}
\usepackage{chemformula}
\usepackage{tabularx}
\usepackage{placeins}
\usepackage{mathtools}
\usepackage{makecell}
\usepackage{array}

\newlength{\smallpic}
\begin{document}

%\title{Efficient implementation of the magnetic force theorem for direct calculation of magnon bands: application to yttrium iron garnet}
\title{Implementation of the magnetic force theorem for large-scale calculations of magnon bands: application to yttrium iron garnet}
%\title{Full magnon spectrum of yttrium iron garnet from first principles calculations} ----Alternativ

\author{Thorbjørn Skovhus} 
\affiliation{CAMD, Computational Atomic-Scale Materials Design, Department of Physics, Technical University of Denmark, 2800 Kgs. Lyngby Denmark}
\author{Varun Rajeev Pavizhakumari} 
\affiliation{CAMD, Computational Atomic-Scale Materials Design, Department of Physics, Technical University of Denmark, 2800 Kgs. Lyngby Denmark}
\author{Thomas Olsen} 
\thanks{Corresponding author. Email: \href{mailto:tolsen@fysik.dtu.dk}{tolsen@fysik.dtu.dk}}
% \email{tolsen@fysik.dtu.dk}
\affiliation{CAMD, Computational Atomic-Scale Materials Design, Department of Physics, Technical University of Denmark, 2800 Kgs. Lyngby Denmark}

\begin{abstract}
We present an efficient implementation of the magnetic force theorem which allows for direct evaluation of exchange parameters in $q$-space. The exchange parameters are calculated directly from Bloch states and the implementation does not rely on any mapping onto localized orbitals. This renders the approach well suited for high-throughput computations, where the construction of a localized basis set---for example Wannier functions---often is impractical.  We demonstrate the versatility of the method by applying it to yttrium iron garnet, where we obtain excellent agreement with the experimental magnon dispersion and Curie temperature without any prior assumptions of important exchange pathways. In particular, the calculations reveal the existence of several inequivalent exchange pathways associated with the same interatomic distances. Performing such calculations in $q$-space fully accounts for long-range exchange interactions %corresponding to a large number of unit cells 
and provides a convenient route for validating models obtained by fitting to inelastic neutron scattering data. %does not have to rely on truncation of real-space exchange paths beyond a certain distance between magnetic atoms. 
\end{abstract}

\maketitle

\section{Introduction}
The fundamental low-energy excitations in magnetically ordered solids---the magnons---constitute a crucial ingredient in the characterization of magnets \cite{Yosida1996}. For example, the magnon energies determine the thermal averages of local magnetic moments as well as the critical temperature for magnetic order \cite{tyablikov1959}. However, the calculation of magnetic excitations from first principles is a profound challenge because of the intricate correlation effects governing the magnons. Nonetheless, \textit{ab initio} magnon spectra have previously been computed using both many-body perturbation theory \cite{Karlsson1999,Kotani_2008,Sasoglu2010,Muller2016,Okumura2019,Friedrich2020,Olsen2021}, time-dependent density functional theory \cite{runge-gross, VanSchilfgaarde1999,Pajda_2001_ab_init_exchange,Buczek2011,Singh2019,Rousseau2012,Cao2018,Skovhus2021, skovhus_minority_2024}, or dynamical mean field theory \cite{Kotliar2006ElectronicStructureWithDMFT, Liechtenstein2001FiniteTemperatureMagnetismDMFT}. Although these approaches have been somewhat successful in terms of reproducing magnon energies in agreement with results from inelastic neutron scattering, such calculations entail severe computational demands and have only been attempted for the simplest magnetic crystals.

Instead of applying direct first principles methods, one may attempt to describe magnons based on models with {\it parameters} obtained from first principles. In this regard, the Heisenberg model plays a central role and has been extensively applied to interpret the magnetic properties of materials. Experimentally, the parameters of the Heisenberg model---the exchange constants---are typically obtained by fitting excitation energies of the model to inelastic neutron scattering data. This approach usually yields excellent agreement between model and measurements and largely corroborates the use of such models to describe the magnetic excitations. In order to predict exchange parameters from first principles, one may apply any total energy method and fit the model to first principles energies associated with different spin configurations. This entails an adiabatic approximation for the spin dynamics \cite{Halilov1998} and may, for example, be carried out in the framework of
density functional theory (DFT) using either constrained spin configurations \cite{Illas1998,Xiang2013, Torelli2019b,Torelli2020}, spin spiral calculations \cite{Bylander1998,Kurz2001,Sandratskii2007,Zimmermann2019,Gutzeit2022,Sodequist_2023} or the magnetic force theorem \cite{LIECHTENSTEIN198765,Antropov1997_exch_inter_in_magnets,Antropov1999Aspects_of_spin_dynamics, Solovyev2021_exch_inter_and_MFT,HE2021107938,Durhuus2023}.

By mapping DFT energies to the Heisenberg model, the accuracy of the resulting magnon energies is mostly limited by the ground state energy functional. In particular, the magnetic force theorem allows one to calculate all exchange parameters without having to resort to super cell calculations or having to neglect spin-orbit coupling as required by spin spiral calculations. The main caveat, however, is the absence of a unique mapping between the itinerant description (obtained from DFT) and the localized spin model used to calculate the magnon energies. Typically, basic intuition may guide one to associate localized spins to particular magnetic atoms in a material, but even that approach is ambiguous since one has to provide a precise definition of what constitutes the atomic sites \cite{Durhuus2023}. In the context of the magnetic force theorem, the most popular choice is to define the spins in terms of orbitals localized at particular sites \cite{LIECHTENSTEIN198765, Korotin2015, HE2021107938}. This can be either atomic orbitals inherent to the applied DFT implementation or Wannier functions \cite{Wannier90}. In either case the result will depend on the applied basis set and whereas it is often (probably correctly) assumed that the basis dependence is rather weak, it is rarely explicitly shown to be the case. Moreover, application of the magnetic force theorem involves a sum over unoccupied states and any result has to be converged with respect to the number of included bands \cite{LIECHTENSTEIN198765,Solovyev2021_exch_inter_and_MFT,Durhuus2023}. This is a rather difficult task for the case of Wannier functions since it is not trivial to add additional bands in the Wannier description. As an alternative to localized orbitals, one may follow a {\it continuum} approach, where the exchange constants are represented in terms of real space integrals of a continuous exchange kernel \cite{Bruno2003,Durhuus2023}. In this case, the sites are rigorously defined by the integrals and do not rely on any localized basis set. However, the exchange kernel is a two-point function and accurate representation quickly becomes very computationally demanding for large super cells. 

In this paper, we extend the applicability of the continuum approach for obtaining exchange constants via the magnetic force theorem by circumventing evaluation of the full two-point exchange kernel. The method maintains the advantage of having well-defined magnetic sites (in real space) and does not rely on localized basis sets. It can be applied seamlessly to any magnetic system for which standard ground state DFT calculations can be carried out. As an example, we apply the method to yttrium iron garnet (YIG), which contains 80 atoms and 20 magnetic sites in the primitive unit cell. The resulting \textit{ab initio} magnon band structure is used to falsify different experimental predictions for the exchange parameters based on (limited) inelastic neutron scattering data.  %We obtain excellent agreement with results obtained from inelastic neutron scattering and 
%we apply the random phase approximations to predict a Curie temperature of XXX K, which is in good agreement with the experimental value of 560 K \cite{CHEREPANOV199381}.

\section{Theory}\label{sec:theory}
The theory underlying adiabatic magnon modeling has been outlined several times in the literature \cite{Halilov1998,LIECHTENSTEIN198765,Katsnelson2004,Durhuus2023}, but for the sake of context, we will briefly review the basic arguments and expressions here. In doing so, we will emphasize the use of real-space domains (rather than localized orbitals) as the basic partitioning used when mapping the magnetic degrees of freedom to a site-based spin model.

\subsection{Magnons from the Heisenberg model}
%To this end, we wish to calculate magnons from the Heisenberg Hamiltonian
%Magnons can generally be described quite accurately with a treatment within the Heisenberg model:
The overarching goal of adiabatic magnon modeling is to calculate the magnetic excitation energies of periodic systems starting from the Heisenberg model: 
\begin{equation}\label{eq:heisenbrghamiltonian}
    H = -\frac{1}{2}\sum_{i,j}\sum_{a,b} J^{ab}_{ij}\, \mathbf{S}^a_{i} \cdot \mathbf{S}^b_{j}.
\end{equation}
The spin operator $\mathbf{S}_i^a$ represents the local magnetic site $a$ in unit cell $i$ and  $J_{ij}^{ab}$ is the exchange interaction between sites $a$ and $b$ in unit cells $i$ and $j$. Here and in the following an isotropic model is assumed, however, the generalization to anisotropic exchange interactions is straightforward.%It is assumed that $J_{ii}^{aa}=0$. 
%The Heisenberg model \eqref{eq:heisenbrghamiltonian} involves the isotropic part of this interaction, but the generalization to anisotropic exchange interactions (driven by spin-orbit coupling) is straightforward and will not be considered in this paper.

%The magnetic excitation spectrum can be derived within the random phase approximation (RPA) for any single-$q$ ordered structure \cite{XXX}. For simplicity we restrict ourselves to isotropic models with equivalent sites, 
Assuming a collinear ground state and applying the random phase approximation (RPA), the magnetic excitation spectrum can be obtained as the positive eigenvalues of the dynamical matrix
\begin{align}
    \textrm{H}^\textrm{RPA}_\mathbf{q} &=
    \begin{bmatrix}
        -A_{\mathbf{q}} + C & -B_{\mathbf{q}} \\
         B_{\mathbf{q}} & A_{\mathbf{q}} - C 
    \end{bmatrix},\label{eq:H_rpa}
\end{align}
where
%With $N_a$ magnetic sites in the unit cell, $\textrm{H}^\textrm{RPA}_\mathbf{q}$ has dimensions $(2N_a \times 2N_a)$ and decomposes into the $(N_a \times N_a)$ matrices
\begin{align}
    A^{ab}_{\mathbf{q}} &= \langle S^a \rangle J^{ab}_{\mathbf{q}}\frac{(1+\mathbf{u}_0^a\cdot\mathbf{u}_0^b)}{2},\\
    B^{ab}_{\mathbf{q}} &= \langle S^a \rangle J^{ab}_{\mathbf{q}}\frac{(1-\mathbf{u}_0^a\cdot\mathbf{u}_0^b)}{2},\\
    C^{ab} &= \delta^{ab}\sum_c \langle S^c \rangle J^{ac}_{\mathbf{0}}\mathbf{u}_0^a\cdot\mathbf{u}_0^c,\label{eq:C}
\end{align}
and
%In order to construct these matrices one needs the ground state spin orientation of the local magnetic sites $\mathbf{u}_0^a$ (unit vectors), the Fourier transform of the exchange constants
%Here $\mathbf{u}_0^a$ (unit vectors) denote the ground state spin orientation of the local magnetic sites,
\begin{align}
    J^{ab}_{\mathbf{q}} &= \sum_i J^{ab}_{0i}  e^{i\mathbf{q}\cdot \mathbf{R}_i}\label{eq:J_q}
\end{align}
is the Fourier transform of the exchange constants. With $N_a$ magnetic sites in the unit cell, $J^{ab}_{\mathbf{q}}$ characterizes the system in terms of $N_a\times N_a$ matrices for each wave vector $\mathbf{q}$ in the first Brillouin zone (BZ). %and $\textrm{H}^\textrm{RPA}_\mathbf{q}$ thus has the dimension $2N_a\times 2N_a$. 
In addition, $\mathbf{u}_0^a$ denotes the direction of the ground state magnetic moment of site $a$ and 
%and the thermal average of the spin operators along $\mathbf{u}_0^a$, 
$\langle S^a \rangle\equiv\langle\mathbf{u}_0^a\cdot \mathbf{S}_i^a \rangle$ is the thermal average of the spin operators along $\mathbf{u}_0^a$. The latter can be calculated from the fluctuation-dissipation theorem \cite{callen1963, Jensen1991} and expressed as 
%In the limit of $T\rightarrow 0$, one may replace $\langle S^a \rangle \rightarrow S^a$, where $S^a$ denotes the maximal eigenvalue of the components of the spin operator $\mathbf{S}_i^a$. As a result, the $T=0$ magnon dispersion can be extracted from a direct diagonalization of the dynamical spin-wave matrix \eqref{eq:H_rpa}. At finite temperatures, however, the thermal average has to be calculated self-consistently. Provided a trial $\langle S^a \rangle$, the dynamical spin-wave matrix is diagonalized, $\textrm{H}^\textrm{RPA}_\mathbf{q} = \textrm{U}_\mathbf{q} \hbar \mathrm{\omega}_\mathbf{q} \textrm{U}_\mathbf{q}^{-1}$, from which a new thermal average can be calculated based on the fluctuation dissipation theorem. In particular,
\begin{equation}\label{eq:magnetization}
    \langle S^a \rangle = \frac{(S^a-\Phi_a)(1+\Phi_a)^{2S^a+1} + (S^a+1+\Phi_a)\Phi_a^{2S^a+1}}{(1+\Phi_a)^{2S^a+1} - \Phi_a^{2S^a+1}},
\end{equation}
where $S^a$ denotes the maximal eigenvalue of the components of the spin operator $\mathbf{S}_i^a$ and
\begin{equation}\label{eq:magnon-number-gfn}
    \Phi_a = \frac{1}{N_q}\sum_{\mathbf{q}} \sum_{\eta=1}^{2N_a}\mathrm{U}_{a \eta \mathbf{q}}\, n_\textrm{B}(\omega_{\eta\mathbf{q}}, T)\, \mathrm{U}^{-1}_{\eta a \mathbf{q}}.
\end{equation}
Here $n_\mathrm{B}$ denotes the Bose distribution at temperature $T$, $N_q$ is the number of $q$-points in the BZ, $\hbar\omega_{\eta\mathbf{q}}$ are the $2 N_a$ eigenvalues associated with %the eigenvector $U_{a\eta\mathbf{q}}$  of 
the dynamical matrix \eqref{eq:H_rpa}, $\textrm{H}^\textrm{RPA}_\mathbf{q} = \textrm{U}_\mathbf{q} \hbar \mathrm{\omega}_\mathbf{q} \textrm{U}_\mathbf{q}^{-1}$, and the $a$ index refers to the first $N_a$ rows of $\textrm{U}$.
%$\omega_{\eta,\mathbf{q}}$ are the eigenvalues of $\textrm{H}^\textrm{RPA}_\mathbf{q}$ and $U_{\mathbf{q}}$ is the matrix that diagonalizes $\textrm{H}^\textrm{RPA}_\mathbf{q}$.

At a given temperature, Eqs. \eqref{eq:H_rpa}-\eqref{eq:magnon-number-gfn} have to be solved self-consistently. Given an initial guess for the magnetization $\langle S^a \rangle$ one may construct and diagonalize the dynamical matrix \eqref{eq:H_rpa}. The magnetization $\langle S^a \rangle$ may then be recalculated from the Brillouin zone average \eqref{eq:magnon-number-gfn} and the whole process can be repeated until a self-consistent solution is reached. If one replaces $\langle S^a \rangle$ by $S^a$ in the dynamical matrix, no self-consistency steps are required and the magnon energies $\hbar\omega_{\eta\mathbf{q}}$ become identical to what is obtained from either Holstein-Primakoff (HP) bosonization or Bloch semiclassical spin-wave theory \cite{rajeev_pavizhakumari_predicting_2025}. In the case of ferromagnets, one has $\langle S^a \rangle=S^a$ for $T=0$ and the three approaches are equivalent. However, for nonferromagnetic systems one has $\langle S^a \rangle(T=0)< S^a$ and the RPA magnon energies are reduced by a factor of $\langle S^a \rangle/S^a$ compared to those obtained from HP or semiclassical spin-wave theory. Similarly, if the exchange parameters of the Heisenberg Hamiltonian \eqref{eq:heisenbrghamiltonian} are fitted to measured magnon energies (at vanishing temperature), one will predict parameters with slightly larger magnitude if fitted to an RPA dispersion compared to that of HP. In any case, one will of course reproduce the experimental dispersion relations as long as one makes a consistent choice (either HP or RPA) for {\it both} the fitting procedure and subsequent calculations. 

%The equations \eqref{eq:H_rpa}-\eqref{eq:magnon-number-gfn} comprise a closed set that must be solved self-consistently to obtain the magnon dispersion $\omega_{\eta,\mathbf{q}}$ and magnetization $\langle S^a \rangle$ at a given temperature. However, in the limit of $T\rightarrow 0$ one may replace $\langle S^a \rangle$ by $S^a$ and the magnon dispersion is readily obtained as the positive eigenvalues obtained from diagonalizing Eq. \eqref{eq:H_rpa}. 
%We also note that the four blocks in $\textrm{H}^\textrm{RPA}_\mathbf{q}$ correspond to raising and lowering of the spin at the different magnetic sites and the eigenstates corresponding to the (physical) positive eigenvalues carry information of the nature (raising or lowering of spin) of the associated magnons.

\subsection{Localized sites and electronic structure theory}
Compared to the full electronic structure problem, the Heisenberg model \eqref{eq:heisenbrghamiltonian} entails a drastic simplification. It implicitly assumes that the low energy magnetic excitations of interest are adiabatically decoupled from higher energy excitations involving interband transitions. 
%only concerns the magnetic degrees of freedom. As such, its adequacy rests on the assumption that the magnetization dynamics of a given material is adiabatically decoupled from its electronic bands. 
Since magnon energies typically reside in the meV range, this assumption is well justified for band insulators. For metals, on the other hand, magnons often couple to inter-band transitions (Stoner excitations) and the adiabatic assumption becomes questionable. %For low-energy magnons, this is a good assumption, but the energy range of its validity will vary between materials. 

%In order to represent the low-energy magnetic properties in real materials in terms of the
In addition to the adiabatic assumption, the Heisenberg model description also requires a choice of magnetic sites for the problem. %In order to map the electronic structure of a given material to the Heisenberg model \eqref{eq:heisenbrghamiltonian}, one first has to define magnetic sites corresponding to the spin operators $\mathbf{S}_i^a$. This is a tricky business because one needs to balance two conflicting considerations: 
%\begin{itemize}
%    \item[(a)] The magnetization of a given site should equal an integer number of Bohr magnetons in order to be properly associated with a quantum mechanical half-integer spin operator.
%    \item[(b)] The internal magnetic degrees freedom of a given site should be fast compared to the collective magnetization dynamics; that is, the magnetization of the entire site should be unidirectional on a magnon time scale.
%\end{itemize}
%The adiabatic assumption (b) favors localized sites restricted to a volume surrounding individual magnetic atoms or clusters of atoms, which in turn violates condition (a) due to the nonzero interstitial magnetization, even for gapped materials. On the other hand, one can for a given site volume $\Omega^a_i$ define a corresponding spin operator
Typically, one choses to include one site for each of the magnetic atoms in the crystal, but one could also choose sites based on clusters of atoms having a collective magnetic moment that is well localized. In any case, it is important to note that there is no unique definition of magnetic sites or local magnetic moments in real materials, only the magnetization {\it density} is well defined. The actual \textit{choice} of site partitioning can be formalized by introducing site volumes $\Omega^a_i$ which in turn define the site-based spin operators \cite{Halilov1998}
\begin{equation}\label{eq:spin_operator}
    \mathbf{S}_i^a = \frac{1}{2}\int_{\Omega^a_i}d\mathbf{r}\, \Psi^\dag(\mathbf{r})\boldsymbol{\sigma}\Psi(\mathbf{r}),
\end{equation}
where $\boldsymbol{\sigma}$ is the Pauli matrix vector and $\Psi(\mathbf{r})$ is the spinorial field operator. If the site volumes $\Omega^a_i$ do not overlap, this definition is easily shown to satisfy the usual spin commutator relations and naturally connects its ground state expectation value,
\begin{equation}\label{eq:spin_operator_av}
    \langle 0|\mathbf{S}_i^a|0\rangle = \frac{1}{2}\int_{\Omega_i^a}d\mathbf{r}\, \mathbf{m}(\mathbf{r}),
\end{equation}
to the ground state magnetization density $\mathbf{m}(\mathbf{r})$ \cite{Durhuus2023}. Thus, if $\mathbf{m}(\mathbf{r})$ is sufficiently well localized within the sites defined by $\Omega^a_i$, the mapping to the site-based model \eqref{eq:heisenbrghamiltonian} becomes insensitive to the exact size and shape of these volumes and can be considered well defined. In principle, the site volumes $\Omega_i^a$ should sum up to the crystal volume and, for insulators, they should be defined such that Eq. \eqref{eq:spin_operator_av} corresponds to the expectation value of a half-integer spin operator. %for a ferromagnetic calculation without spin-orbit coupling. 
However, this is typically an overly restrictive requirement, since the magnon energies generally are governed by the magnetization localized at the magnetic atoms, meaning that the small remnant of magnetization at ligands and interstitial regions can be effectively neglected. 
In addition, for metals, the localized spins are not expected to be half-integer, but one may nonetheless still formally define quantum spin operators based on Eq. \eqref{eq:spin_operator}. However, due to the presence of longitudinal fluctuations, it is often sufficient to treat metals in a classical limit of the Heisenberg model \eqref{eq:heisenbrghamiltonian}.% and simply take the classical spin vectors to have arbitrary---yet constant---length.

\subsection{Exchange interactions from DFT}
%In conjunction with the spin-operator definition \eqref{eq:spin_operator}, the mapping from electronic structure theory to Heisenberg model is completed by the Liechenstein formula for the exchange constants $J^{ab}_{ij}$ \cite{LIECHTENSTEIN198765}. 

In order to make contact between the electronic structure and the localized spin model, one considers the expectation value of the Heisenberg Hamiltonian \eqref{eq:heisenbrghamiltonian} with respect to states $|\{\mathbf{u}^a_i\}\rangle$ where the spins have been aligned along a set of unit vectors $\{\mathbf{u}^a_i\}$. %That is $\mathbf{u}^a_i\cdot\mathbf{S}^a_i|\{\mathbf{u}^a_i\}\rangle=S^a|\{\mathbf{u}^a_i\}\rangle$. and $S^a$ denotes the maximal eigenvalue of a component of the spin operator $\mathbf{S}_i^a$. We then have
This can be expressed as
\begin{subequations}\label{eq:heisenbrghamiltonian_av}
\begin{align}
    \langle H\rangle &= -\frac{1}{2}\sum_{ijab}{}^{'} J^{ab}_{ij}\langle\mathbf{S}^a_{i}\rangle\cdot \langle\mathbf{S}^b_{j}\rangle+\tilde E\\
    &= -\frac{1}{2}\sum_{ijab}{}^{'} \tilde J^{ab}_{ij}\mathbf{u}^a_i\cdot\mathbf{u}^b_j+\tilde E,
\end{align}
\end{subequations}
where $\sum{}^{'}$ implies a summation which excludes $a=b$ if $i=j$, 
\begin{equation}
    \tilde E=\frac{N_a}{2}\sum_a J_{00}^{aa}S^a(S^a+1),
\end{equation}
and 
\begin{align}\label{eq:Jt}
    \tilde J^{ab}_{ij}\equiv S^aS^bJ^{ab}_{ij}.
\end{align}
Since the expectation value coincides with the corresponding classical energy of the spin configuration $\{\mathbf{u}^a_i\}$ (up to a constant), one may regard variations in $\langle H\rangle$ as a simple function of the classical spin configuration defined by $\{\mathbf{u}^a_i\}$.% and an implicit functional of the classical field $\mathbf{m}(\mathbf{r})$. 

%Consider now the DFT total energy functional $E[n,\mathbf{m}]$ which yields the energy of a system of electrons with ground state density $n$ and magnetization $\mathbf{m}$, subject to a fixed (material-specific) nuclei potential. 
%In the framework of DFT 
In the framework of DFT, the ground state energy is typically obtained by minimizing the total energy $E[n,\mathbf{m}]$ regarded as a functional of the density $n$ and magnetization $\mathbf{m}$. Alternatively, one may define an adiabatic energy functional $E^\mathrm{A}[\mathbf{u}]$, which minimizes $E[n,\mathbf{m}]$ under a fixed {\it direction} of magnetization $\mathbf{u}$ and yields the correct ground state $\mathbf{u}_0$  when minimized with respect to $\mathbf{u}$. % thanks to the Hohenberg-Kohn theorems \cite{HohenbergP.1973}. %It is easy to see that the ground state energy can be obtained by minimizing this, since one may minimize $n$ and $m=|\mathbf{m}|$ for a given function $\mathbf{u}$ and the minimization with respect to $\mathbf{u}$ is thus obtained as a two-step process. 
For small deviations in $\mathbf{u}$ with respect to $\mathbf{u}_0$, the adiabatic energy functional may be written as a quadratic expansion in the classical field $\mathbf{u}(\mathbf{r})$. Neglecting spin-orbit effects, % we can then consider the quadratic expansion
\begin{equation}\label{eq:adiabatic_energy}
    E^\mathrm{A}_0[\mathbf{u}] = E_0[\mathbf{u}_0] + \frac{1}{2}\iint d\mathbf{r}d\mathbf{r}'A(\mathbf{r},\mathbf{r}')\delta\mathbf{u}(\mathbf{r})\cdot\delta\mathbf{u}(\mathbf{r}'),
\end{equation}
where $\delta\mathbf{u}(\mathbf{r})=\mathbf{u}(\mathbf{r})-\mathbf{u}_0(\mathbf{r})$. 
%As a result, the exchange constants $J^{ab}_{ij}$ of a given material can be uniquely defined given a specific site partitioning \eqref{eq:spin_operator} by requiring that the classical Heisenberg energy \eqref{eq:heisenbrghamiltonian_av} varies in an identical fashion to the adiabatic energy functional \eqref{eq:adiabatic_energy} as a function of infinitesimal rotations of the site magnetization (with respect to the ground state):As a result of the obvious similarity between Eqs. \eqref{eq:heisenbrghamiltonian_av} and \eqref{eq:adiabatic_energy}, a unique definition for $J^{ab}_{ij}$ can be established. Namely, given a specific site partitioning \eqref{eq:spin_operator}, one considers infinitesimal rotations of the site magnetization (with respect to the ground state),
%In order to make contact to the localized model \eqref{eq:heisenbrghamiltonian} we 
The final step is to consider variations in the magnetization direction which can be written in terms of a discrete set of unit vectors such that
\begin{align}
\delta\mathbf{u}(\mathbf{r})= \sum_i\sum_a\Theta(\mathbf{r} \in \Omega^a_i)\delta\mathbf{u}^a_i,
\label{eq:rotating the site magnetization}
\end{align}
where $\Theta(\mathbf{r} \in \Omega^a_i)$ denotes a step function which equals unity inside the site volume $\Omega^a_i$ and vanishes outside it.  Comparison with Eq.  \eqref{eq:heisenbrghamiltonian_av} then uniquely determines $J^{ab}_{ij}$ in terms of $A(\mathbf{r}, \mathbf{r}')$.% varies in an identical fashion to the adiabatic energy functional \eqref{eq:adiabatic_energy} as a function of infinitesimal rotations \eqref{eq:rotating the site magnetization}.
%Inserting this into Eq. \eqref{eq:adiabatic_energy} and comparing with Eq. \eqref{eq:heisenbrghamiltonian_av} we get
%\begin{align}\label{eq:Jt_ij}
%\tilde J^{ab}_{ij}=-\int_{\Omega^a_i}\int_{\Omega^b_j}d\mathbf{r}d\mathbf{r}'A(\mathbf{r}, \mathbf{r}'),
%\end{align}
%for $i\neq j$ if $a=b$, 

In principle, the kernel $A(\mathbf{r}, \mathbf{r}')$ can be obtained from DFT energy differences by realizing small rotations of the site magnetization \eqref{eq:rotating the site magnetization} via a constraint. This entails a full mapping between the adiabatic energy functional and the Heisenberg model. However, one should bear in mind that the expectation values in Eq. \eqref{eq:heisenbrghamiltonian_av} are with respect to a state of rigidly rotated spin directions, whereas the adiabatic functional \eqref{eq:adiabatic_energy} implies that longitudinal degrees of freedom are relaxed for a given field configuration $\mathbf{u}(\mathbf{r})$. For insulators, the magnitude of the magnetization may be assumed constant if the rotations are sufficiently small, and in that case the approach implies a direct mapping to the {\it quantum mechanical Heisenberg model} \eqref{eq:heisenbrghamiltonian_av}. For metals, such an assumption is not justified and a quantum mechanical Heisenberg description becomes dubious. Nevertheless, in a {\it classical} description, one may regard the energy as a function of $\mathbf{u}_i$ without reference to the magnitude of spin. This implies a well-defined mapping of DFT energies to a classical Heisenberg model, where $\tilde J^{ab}_{ij}$ rather than $J^{ab}_{ij}$ constitute the fundamental exchange parameters. % On the other hand, the expansion \eqref{eq:adiabatic_energy} is only valid for infinitesimal deviations from the ground state. For such states, it is reasonable to assume vanishing effects of self-consistency and 
%However, given that the rotations \eqref{eq:rotating the site magnetization} need to be small for the quadratic expansion \eqref{eq:adiabatic_energy} to hold, 

Instead of realizing the spin rotations \eqref{eq:rotating the site magnetization} in actual simulations, one can make use of the magnetic force theorem which implies that the change in energy is contained solely in the change in Kohn-Sham eigenenergies. This change may be evaluated from second-order perturbation theory, resulting in the isotropic exchange interaction \cite{Durhuus2023}
\begin{align}\label{eq:Liechtenstein}
\tilde J^{ab}_{ij}=-2\int_{\Omega^a_i}\int_{\Omega^b_j}d\mathbf{r}d\mathbf{r}'\, B^{\mathrm{xc}}(\mathbf{r}) \chi_{\mathrm{KS}}'^{+-}(\mathbf{r}, \mathbf{r}') B^{\mathrm{xc}}(\mathbf{r}').
\end{align}
Here $B^{\mathrm{xc}}(\mathbf{r})$ is the functional derivative of the exchange-correlation energy with respect to the magnitude of $\mathbf{m}(\mathbf{r})$ and
\begin{align}
    \chi'^{+-}_{\mathrm{KS}}(\mathbf{r},\mathbf{r}') = & \sum_{n,m} \frac{f_{n\uparrow} - f_{m\downarrow}}{\varepsilon_{n\uparrow} - \varepsilon_{m\downarrow}} \psi_{n\uparrow}^*(\mathbf{r}) \psi_{m\downarrow}(\mathbf{r}) 
    \psi_{m\downarrow}^*(\mathbf{r}')
    \psi_{n\uparrow}(\mathbf{r}')
\end{align}
is the reactive part of the static transverse Kohn-Sham susceptibility. %It should be noted that Eq. \eqref{eq:Liechtenstein} is derived excluding onsite exchange interactions. 
%As a final remark, one may note that onsite exchange interactions have not been left out of the Heisenberg model \eqref{eq:heisenbrghamiltonian}, as is common practice in literature. Now, since $J_{ii}^{aa}$ only provides a rigid energy shift of all expectation values, it is a free variable rather than a physical quantity, and one can therefore take Eq. \eqref{eq:Liechtenstein} to define the onsite interaction as well as the physical exchange interactions, without any loss of generality.
 
\subsection{Calculating exchange interactions efficiently}

%In this paper, a novel implementation of the Liechenstein formula \eqref{eq:Liechtenstein} is presented. The central idea of the implementation is to leverage the direct relation between exchange constants and Kohn-Sham susceptibility to recast the Liechenstein formula \eqref{eq:Liechtenstein} as the static limit of a generalized linear response function. 
The evaluation of Eq. \eqref{eq:Liechtenstein} is typically handled by representing the susceptibility in a plane-wave basis \cite{Durhuus2023} or in a localized basis set \cite{HE2021107938}. While the former approach becomes prohibitively expensive for large systems, the latter approach can be difficult to converge with respect to unoccupied bands and basis functions. However, instead of calculating the full two-point susceptibility $\chi'^{+-}_{\mathrm{KS}}(\mathbf{r},\mathbf{r}')$, we propose to evaluate the integrals% The novelty of this paper is the way we approach the calculation of the exchange constants \eqref{eq:Liechtenstein} in practice. Normally, one would first calculate the Kohn-Sham susceptibility $\chi'^{+-}_{\mathrm{KS}}(\mathbf{r},\mathbf{r}')$ \cite{Durhuus2023} or the corresponding single-particle Green's functions $G^\uparrow_{\mathrm{KS}}(\mathbf{r},\mathbf{r}', \omega)$ and $G^\downarrow_{\mathrm{KS}}(\mathbf{r},\mathbf{r}', \omega)$ \cite{Bruno2003} in a real-space basis of choice and subsequently evaluate the product \eqref{eq:Liechtenstein} with $B^\mathrm{xc}(\mathbf{r})$. The problem with this approach is the basis function dependency. Either one chooses a basis set such as plane waves which can be systematically expanded to reach completeness (computationally expensive) or a noncomplete localized basis set (typically Wannier functions) which is optimized for each specific material, but in a way which is hard to generalize.
%Instead, we propose to calculate the matrix elements %site spin pair energies,
%\begin{subequations}
\begin{align}\label{eq:site spin pair energy}
    d^{\mathrm{xc},ai}_{ns,ms'} %&\equiv 
    %\langle \psi_{ns}| \Theta(\mathbf{r} \in \Omega^a_i)\, %B^{\mathrm{xc}}(\mathbf{r}) |\psi_{ms'}\rangle \\
    &= \int_{\Omega^a_i} d\mathbf{r}\, \psi_{ns}^*(\mathbf{r})  B^{\mathrm{xc}}(\mathbf{r}) \psi_{ms'}(\mathbf{r}),
\end{align}
%\end{subequations}
from which the exchange constants can be calculated as a straightforward sum over states:
\begin{equation}\label{eq:exchange as linear response}
    \tilde J^{ab}_{ij} = - 2 \sum_{n,m} \frac{f_{n\uparrow} - f_{m\downarrow}}{\varepsilon_{n\uparrow} - \varepsilon_{m\downarrow}} d^{\mathrm{xc},ai}_{n\uparrow,m\downarrow} d^{\mathrm{xc},bj}_{m\downarrow,n\uparrow}.
\end{equation}
% In this formulation, the exchange constants are directly related to the reactive part of the linear response function
% \begin{equation}
%     \iota_{ij}^{ab}(\omega) = \lim_{\eta\rightarrow 0^+} \sum_{n,m} \frac{f_{n\uparrow} - f_{m\downarrow}}{\hbar\omega - (\varepsilon_{m\downarrow}-\varepsilon_{n\uparrow}) + i\hbar\eta} d^{\mathrm{xc},ai}_{n\uparrow,m\downarrow} d^{\mathrm{xc},bj}_{m\downarrow,n\uparrow},
% \end{equation}
% with $\tilde J^{ab}_{ij}=-2\iota'^{ab}_{ij}(\omega=0)$.
%This is an attractive approach since one can use the preexisting representation of the Kohn-Sham orbitals $\psi_{ns}(\mathbf{r})$ of a given DFT code to evaluate the site spin pair energies \eqref{eq:site spin pair energy} precisely, thus circumventing the aforementioned basis function dependency.
This approach has the major advantage that the integral \eqref{eq:site spin pair energy} involves only products of one-point functions, namely $\psi_{ns}$ and $B^\mathrm{xc}$. As a result, the exchange constants are readily calculated for any system where an ordinary DFT calculation is feasible, and convergence with respect to parameters such as plane-wave cutoff is straightforward.

%\subsection{Exchange constants}
%In GPAW, the Kohn-Sham orbitals of 
In periodic systems, computations may be simplified further by calculating the exchange constants in reciprocal space. Specifically, the Kohn-Sham orbitals can be expressed as Bloch waves, $\psi_{ns}(\mathbf{r})\rightarrow\psi_{n\mathbf{k}s}(\mathbf{r})/\sqrt{N_k}$, where $\psi_{n\mathbf{k}s}(\mathbf{r}) = e^{i\mathbf{k}\cdot\mathbf{r}} u_{n\mathbf{k}s}(\mathbf{r})$ is normalized to the unit cell and $u_{n\mathbf{k}s}(\mathbf{r}+\mathbf{R}_i) = u_{n\mathbf{k}s}(\mathbf{r})$ is periodic under translation of a lattice vector $\mathbf{R}_i$. With a grid of $N_k$ $k$-points (corresponding to $N_k$ unit cells in the crystal), the reactive part of the static Kohn-Sham susceptibility is given by
\begin{align}
    \chi'^{+-}_{\mathrm{KS}}(\mathbf{r},\mathbf{r}') = &\frac{1}{N_k^2} \sum_{\mathbf{k},\mathbf{k}'}\sum_{n,m} \frac{f_{n\mathbf{k}\uparrow} - f_{m\mathbf{k}'\downarrow}}{\varepsilon_{n\mathbf{k}\uparrow} - \varepsilon_{m\mathbf{k}'\downarrow}} \label{eq:chi^+-_KS}\\
    &\times\psi_{n\mathbf{k}\uparrow}^*(\mathbf{r}) \psi_{m\mathbf{k}'\downarrow}(\mathbf{r}) 
    \psi_{m\mathbf{k}'\downarrow}^*(\mathbf{r}')
    \psi_{n\mathbf{k}\uparrow}(\mathbf{r}'). \notag
\end{align}
Thanks to the periodicity of the exchange-correlation magnetic field $B^\mathrm{xc}(\mathbf{r}+\mathbf{R}_i)=B^\mathrm{xc}(\mathbf{r})$, the Fourier transformed exchange constants \eqref{eq:J_q} only involves transitions that satisfy $\mathbf{k}'-\mathbf{k}=\mathbf{q}$ modulo a reciprocal lattice vector. In particular,
%Here $\psi_{n\mathbf{k}s}(\mathbf{r})$ denotes the $n$'th Kohn-Sham eigenfunction at $k$-point $\mathbf{k}$ and spin channel $s$ with eigenvalue $\varepsilon_{n\mathbf{k}s}$ and occupation number $f_{n\mathbf{k}s}$. Rather than evaluating the response function explicitly, we may insert Eqs. \eqref{eq:A_rr} and \eqref{eq:chi^+-_KS} in Eq. \eqref{eq:Jt_ij} and get
% \begin{align}
%     \tilde J^{ab}_{ij} = -2\sum_{\mathbf{k}\mathbf{q}\in\mathrm{BZ}}\sum_{nm} &\frac{f_{n\mathbf{k}\uparrow} - f_{m\mathbf{k}+\mathbf{q}\downarrow}}{\varepsilon_{n\mathbf{k}\uparrow} - \varepsilon_{m\mathbf{k}+\mathbf{q}\downarrow}} d^a_{n\mathbf{k}m\mathbf{k}+\mathbf{q},i}d^b_{m\mathbf{k}+\mathbf{q}n\mathbf{k},j} \label{eq:Jt_divided}
% \end{align}
% with
% \begin{align}
%  d^a_{n\mathbf{k}m\mathbf{k}+\mathbf{q},i}=&\label{eq:d_nm}\int_{\Omega^a_i}d\mathbf{r}\psi_{n\mathbf{k}\uparrow}^*(\mathbf{r}) \psi_{m\mathbf{k}+\mathbf{q}\downarrow}(\mathbf{r})B^{\mathrm{xc}}(\mathbf{r})\notag\\
%  =&e^{i\mathbf{q}\cdot\mathbf{R}_i}\int_{\Omega^a_0}d\mathbf{r}\psi_{n\mathbf{k}\uparrow}^*(\mathbf{r}) \psi_{m\mathbf{k}+\mathbf{q}\downarrow}(\mathbf{r})B^{\mathrm{xc}}(\mathbf{r})\notag\\
%  \equiv&e^{i\mathbf{q}\cdot\mathbf{R}_i}d^a_{n\mathbf{k}m\mathbf{k}+\mathbf{q}}.
% \end{align}
% where $\mathbf{R}_i$ is the lattice vector corresponding to unit cell $i$. Using Eq. \eqref{eq:J_q} it is then straightforward to obtain
\begin{align}
    \tilde{J}^{ab}_\mathbf{q}= -\frac{2}{N_k}\sum_{\mathbf{k}}\sum_{n,m} &\frac{f_{n\mathbf{k}\uparrow} - f_{m\mathbf{k}+\mathbf{q}\downarrow}}{\varepsilon_{n\mathbf{k}\uparrow} - \varepsilon_{m\mathbf{k}+\mathbf{q}\downarrow}} d^{\mathrm{xc},a}_{n\mathbf{k}\uparrow,m\mathbf{k}+\mathbf{q}\downarrow}d^{\mathrm{xc},b}_{m\mathbf{k}+\mathbf{q}\downarrow,n\mathbf{k}\uparrow}, \label{eq:Jt_q}
\end{align}
%We note that in principle one has to subtract $\delta^{ab}J_{00}^{aa}$ from this expression since we excluded the $i=j$ terms for $a=b$ in Eq. \eqref{eq:Jt_ij}. However, such a term is irrelevant for magnon spectra since it cancels out in Eqs. \eqref{eq:H_rpa}-\eqref{eq:C} and we disregard it here.
%In order to calculate $\tilde{J}^{ab}_\mathbf{q}$ for a given wave vector $\mathbf{q}$---from which the magnon bands can be inferred according to Eqs. \eqref{eq:H_rpa} and \eqref{eq:rpa Na x Na matrices} using also the length of the spin operators $S^a$ and $\tilde{J}^{ab}_\mathbf{0}$---all that one needs in addition to the Kohn-Sham eigenvalues and occupations are therefore the site spin pair energies evaluated in the first unit cell,
with site spin pair energies given by
\begin{equation}\label{eq:dxc nks}
     d^{\mathrm{xc},a}_{n\mathbf{k}s,m\mathbf{k}'s'} = \int_{\Omega^a}d\mathbf{r}\, \psi^*_{n\mathbf{k}s}(\mathbf{r})B^{\mathrm{xc}}(\mathbf{r})\psi_{m\mathbf{k}'s'}(\mathbf{r}).
    %\langle \psi_{n\mathbf{k}s}| \Theta(\mathbf{r} \in \Omega_a)\, B^{\mathrm{xc}}(\mathbf{r}) |\psi_{m\mathbf{k}'s'}\rangle.
\end{equation}
%for all pairs of wave vectors $\mathbf{k}$ and $\mathbf{k}'$ connected by $\mathbf{q}$. 
%and $\mathbf{k}+\mathbf{q}$ referring to the first BZ $\mathbf{k}'=\mathbf{k}+\mathbf{q} + \mathbf{G}$. 
%Left is only the DFT code specific implementation of the matrix elements \eqref{eq:dxc nks}. 
Since the RPA Hamiltonian \eqref{eq:H_rpa} depends explicitly on $\tilde{J}^{ab}_\mathbf{q}$, Eqs. \eqref{eq:Jt_q}-\eqref{eq:dxc nks} provide a highly convenient starting point for calculating magnon energies. In Appendix \ref{sec:gpaw implementation} we present an implementation of Eq. \eqref{eq:dxc nks} in the electronic structure code GPAW \cite{Mortensen2024} using the projector augmented wave method.

\section{Application to YIG}

\begin{figure*}[tb]
     \centering
     \includegraphics[scale=1.0]{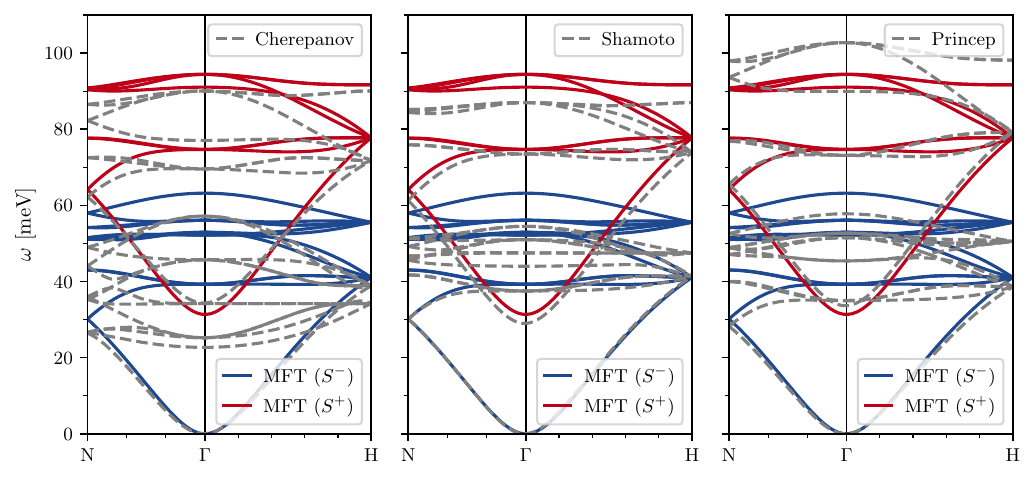}
     \caption{Magnon spectrum of YIG calculated with LDA+U and U = 6.0 eV. The eigenvalues have been colored according to whether the eigenstate is lowering spins in the majority channel ($S^-$) or raising spins in the minority channel ($S^+$). Left: comparison to an INS data fitted model from Cherepanov et al. \cite{CHEREPANOV199381}. Center: comparison to an INS data fitted model from Shamoto et al. \cite{Shamoto2018}. Right: comparison to an INS data fitted model from Princep et al. \cite{Princep2017}.}
     \label{fig:YIG_magnons}
\end{figure*}
Yttrium iron garnet (YIG) has become the material of choice for investigating magnon transport, primarily due to the long spin relaxation time and low damping rate of magnons \cite{yig_review}. It crystallizes in the body-centered cubic lattice with four units of the basic formula Y$_3$Fe$_5$O$_{12}$ in the primitive cell. The structure has been discussed in detail elsewhere \cite{10.1063/1.5051045,Shamoto2018} and here we just note that the Fe atoms carry magnetic moments of 5 $\mu_\mathrm{B}$ and that the magnetic ground state is ferrimagnetic with 12 tetrahedrally coordinated sites composing the majority spin channel and 8 octahedrally coordinated sites having minority spins. The 20 magnetic atoms in the primitive cell imply 20 magnon bands, but the far majority of magnonics applications only involve the low-energy acoustic magnon in vicinity of the Brillouin zone center. However, due to the importance of YIG for technological applications, extensive efforts have been made to determine the full magnon spectrum and exchange parameters from inelastic neutron scattering (INS) experiments \cite{CHEREPANOV199381,Princep2017,Shamoto2018}. Although the acoustic magnon is well accounted for in most fits to experimental data, there is still considerable uncertainty in the exact position of optical magnon energies. Moreover, the large number of possibly non-vanishing exchange parameters may permit several qualitatively different fits that match the experimental data equally well. For example, Princep et al. noted that there are two different exchange pathways for the important third nearest neighbor interaction and allowing the associated exchange constants to differ, changes the entire set of fitted exchange constants significantly \cite{Princep2017}. To this end, first principles calculations may prove useful since such approaches do not come with implicit assumptions of dominating or negligible exchange pathways. In Ref. \cite{PhysRevB.95.014423} the exchange constants were calculated from an energy mapping approach using super cells and in Ref. \cite{PhysRevB.104.174401} the magnon dispersion was calculated using the magnetic force theorem. In both cases, reasonable agreement was obtained with the measured acoustic magnons, but the inequivalent exchange paths noted by Princep et al. \cite{Princep2017} were not taken into account.%In both cases the results were in reasonable agreement with a recent fit to experimental data by Shamoto et al. \cite{Shamoto2018}, but the optical modes disagreed with a different fit reported by Cherepanov et al. \cite{CHEREPANOV199381}.

\begin{table*}[t]
    \centering
    \begin{tabular}{
        | >{\centering\arraybackslash}p{2cm} 
        | >{\centering\arraybackslash}p{2cm} 
        | >{\centering\arraybackslash}p{2cm} 
        | >{\centering\arraybackslash}p{2cm} 
        | >{\centering\arraybackslash}p{2cm} 
        | >{\centering\arraybackslash}p{2cm} 
        | >{\centering\arraybackslash}p{2cm} |
    }
        \hline
        &&&&&&\\
        $\Delta R_n(\mathrm{\AA})$ & $J_n$ (meV) & $N_n$ &  Ref.~\cite{CHEREPANOV199381} & Ref.~\cite{Shamoto2018} & Ref.~\cite{Princep2017} & This work \\
        &&&&&&\\
        \hline
        &&&&&&\\
        3.459& $J_1^{ot}$ & 6/4 & -6.86 & -5.80 & -6.8 & -6.211 \\
        &&&&&&\\
        3.789& $J_2^{tt}$ & 4 & -2.31 & -0.70 & -0.52 & -0.004 \\
        &&&&&&\\
        5.359 & \makecell[c]{$J_{3}^{oo,1}$ \\ $J_{3}^{oo,2}$}
              & \makecell[c]{6 \\ 2}
              & \makecell[c]{-0.65 \\ -0.65}
              & \makecell[c]{ \\ }
              & \makecell[c]{0.0 \\ -1.1}
              & \makecell[c]{0.051 \\ -0.450} \\
        &&&&&&\\
        5.578 & $J_4^{ot}$ & 6/4 &  &  & 0.07 & -0.043 \\
        &&&&&&\\
        5.788 & $J_5^{tt}$ & 8 & &  & -0.47 & -0.347 \\
        &&&&&&\\
        6.188 & \makecell[c]{$J_{6}^{oo}$ \\ $J_{6}^{tt}$}
              & \makecell[c]{6 \\ 2}
              & \makecell[c]{ \\ }
              & \makecell[c]{ \\ }
              & \makecell[c]{-0.09 \\ }
              & \makecell[c]{-0.029 \\ -0.074} \\
        &&&&&&\\
        6.918 & \makecell[c]{$J_{7}^{tt,1}$ \\ $J_{7}^{tt,2}$}
              & \makecell[c]{4 \\ 4}
              & \makecell[c]{\\}
              & \makecell[c]{\\}
              & \makecell[c]{\\}
              & \makecell[c]{-0.050 \\ 0.069} \\
        &&&&&&\\
        7.089 & \makecell[c]{$J_{8}^{ot,1}$ \\ $J_{8}^{ot,2}$}
              & \makecell[c]{6/4 \\ 6/4}
              & \makecell[c]{\\}
              & \makecell[c]{\\}
              & \makecell[c]{\\}
              & \makecell[c]{-0.002 \\ -0.012} \\
        &&&&&&\\
        7.256 & $J_9^{tt}$ & 4 &  &  &  & -0.004 \\
        &&&&&&\\
        8.331 & \makecell[c]{$J_{10}^{ot,1}$ \\ $J_{10}^{ot,2}$ \\$J_{10}^{ot,3}$}
              & \makecell[c]{6/4 \\ 6/4 \\ 6/4}
              & \makecell[c]{\\ \\}
              & \makecell[c]{\\ \\}
              & \makecell[c]{\\ \\}
              & \makecell[c]{-0.001 \\ 0.0 \\ 0.0} \\
        &&&&&&\\
        8.473 & $J_{11}^{tt}$ & 8 &  &  &  & 0.001 \\
        &&&&&&\\
        \hline
        \multicolumn{3}{|c|}{}&&&&\\
        \multicolumn{3}{|c|}{$T_\mathrm{C}^\mathrm{RPA}$ (K)} & 480 & 552 & 540 & 570 \\
        \multicolumn{3}{|c|}{}&&&&\\
        \hline
    \end{tabular}
    \caption{Exchange constants and RPA critical temperature of YIG. The $t$ and $o$ superscripts denote tetrahedral and octahedral sites, respectively, $\Delta R_n$ is the $n^{th}$ nearest-neighbor distance and $N_n$ denotes the number of nearest-neighbor sites at this distance (fractions indicate number of tetragonal/octahedral neighbors). In several cases there are inequivalent exchange paths at the same distance. For example, the exchange pathway for $J_{3}^{oo,1}$ has $C_2$ symmetry, whereas $J_{3}^{oo,2}$ has $D_3$ symmetry \cite{Princep2017}.}
    \label{table:J_n}
\end{table*}
In the following, we use YIG to exemplify the site-based evaluation of exchange constants as implemented in the electronic structure code GPAW \cite{Mortensen2024}. The structure was taken from Ref. \cite{10.1063/1.5051045}, and the collinear ground state calculation was performed using a plane-wave cutoff of 800 eV and a $4\times4\times4$ uniform $k$-point grid. With the default PAW setups in GPAW this yields a structure with 580 valence electrons. We used the LDA+U functional with an effective U parameter as proposed by Dudarev et al. \cite{PhysRevB.57.1505}, and included a total of 400 bands in each spin channel for the calculation of the exchange constants. In Appendix \ref{sec:mft_U} we document the inclusion of Hubbard corrections in the MFT calculations. The total magnetic moment obtained from the DFT ground state is 20 $\mu_\mathrm{B}$ (12 majority and 8 minority sites), in support of a value of $S^a=5/2$ for all Fe atoms. The magnon dispersion obtained with $\mathrm{U}=6.0$ eV and $S=5/2$ is shown in Fig. \ref{fig:YIG_magnons} together with the experimental dispersions reported by  Cherepanov et al. \cite{CHEREPANOV199381}, Shamoto et al. \cite{Shamoto2018} and Princep et al. \cite{Princep2017}. The calculated magnon energies are seen to be in qualitative agreement with the Shamoto dispersion, but deviate qualitatively from both the Cherepanov results and the Princep results. Most prominently, one may note the appearance of an isolated band just below 80 meV at the $\Gamma$-point in the Cherepanov dispersion (residing around 90 meV in the Princep dispersion), which is absent in the calculated spectrum and the Shamoto fit. Moreover, our calculations predict the highly dispersive $S^+$-band to reside well below the optical $S^-$ bands at the $\Gamma$-point, which is in excellent agreement with the Shamoto fit, but not with the Cherepanov or Princep fits. We note that the magnon band width is rather sensitive to the choice of U, which introduces an overall scaling of the magnon dispersion. Increasing U reduces magnon energies and the value of 6.0 eV was chosen to reproduce the acoustic Goldstone branch obtained from the exchange constants of Shamoto et al. %However, even if we choose U to agree with the band width of Cherepanov, the best overall agreement is with the Shamoto results. 
%We also note that this choice of U yields excellent agreement of the entire acoustic magnon branch (which is directly measured) with the Cherepanov et al. and Princep et al. fits.

The expression \eqref{eq:Jt_q} yields the exchange constants in $q$-space from which the magnon energies may be calculated directly. In order to obtain the exchange constants in real space, we have calculated $\tilde J^{ab}(\mathbf{q})$ on a uniform $4\times4\times4$ $q$-point grid and Fourier transformed the results. The resulting exchange constants are tabulated in Tab. \eqref{table:J_n} and compared with the fitted values from Cherepanov \cite{CHEREPANOV199381}, Shamoto \cite{Shamoto2018} and Princep \cite{Princep2017}. By using a $4\times4\times4$ grid, {\it all} exchange constants within a distance of four unit cells are obtained. At a given distance, we automatically get all inequivalent exchange constants and we do not have to explicitly determine whether or not paths are equivalent by symmetry or not. In particular, we find that for the third nearest neighbor exchange, the high symmetry path $J_3^{oo,2}$ is larger than the low symmetry path $J_3^{oo,1}$ similar to the fitted values of Princep et al. \cite{Princep2017}. Moreover, we find two inequivalent paths for the 6'th, 7'th and 8'th nearest neighbors and three inequivalent paths for the 10'th nearest neighbors. We state all inequivalent interactions up to a distance of 8.5 {\AA} in Tab. \ref{table:J_n} and find that the magnitude of all interactions beyond this point is below 1 $\mu$eV. We note that all interactions beyond $J_7$ have minor influence on the magnon spectrum and it is probably safe to ignore these. However, it is important to stress that this is a result of the present analysis where the evaluation in $q$-space allow us to extract {\it all} exchange constants and the decay of exchange beyond $J_7$ is {\it observed}. This is in sharp contrast to real space implementations of MFT where the real space exchange constants under scrutiny are {\it chosen} prior to a calculation and the user has to know about inequivalent paths to capture inequivalent exchange constants at a given distance \cite{PhysRevB.104.174401}.

In Tab. \ref{table:J_n} we also state the Curie temperatures calculated from RPA using the exchange constants from the three experimental references as well as the ones calculated here. The numbers should be compared to the experimental value of 559 K \cite{CHEREPANOV199381} and the largest deviation is that of Cherapanov et al, which underestimates it by 79 K. %This is likely due to a general overestimation of the majority magnon band energies compared to Princep et al. and the present work. 
We note that the experimental exchange constants are obtained from the Holstein-Primakoff dispersion relation, which is related to RPA by taking $\langle S^a\rangle\rightarrow S^a$ in Eqs. \eqref{eq:H_rpa}-\eqref{eq:C}. For ferromagnets $\langle S^a\rangle=S^a$ at $T=0$, but for the case of YIG $\langle S^a\rangle/S^a=0.95(0.97)$ for octahedral(tetrahedral) sites, which implies that a fit to the RPA dispersion would yield larger exchange constants (by a factor of $S^a/\langle S^a\rangle$). In Fig. \ref{fig:YIG_magnons} and Tab. \ref{table:J_n}, we therefore present the Holstein-Primakoff $T=0$ magnon dispersion and exchange constants in order to make the comparison to experimental references straightforward. However, this implies that the application of RPA for calculating critical temperatures involves a small inconsistency, which leads to slightly underestimated critical temperatures. Nevertheless, RPA is still expected to perform significantly better compared to either classical Monte Carlo simulations or Holstein-Primakoff 
critical temperatures \cite{rajeev_pavizhakumari_predicting_2025}.

\section{Discussion}

In addition to the quality of the exchange-correlation functional, the accuracy of the present approach is fundamentally limited by the extent to which the problem may be addressed based on a site-based model in the first place. For the YIG calculations presented above we used spherical sites $\Omega^a_i$ centered on the Fe atoms with a radius of 1.165 {\AA} to calculate rescaled exchange constants $\tilde{J}^{ab}_\mathbf{q}$. The definition of sites is somewhat arbitrary, but we may regard the mapping to the site model as being well defined if the results are largely insensitive to the choice of sites. 
\begin{figure}[tb]
     \centering
     \includegraphics[scale=1.0]{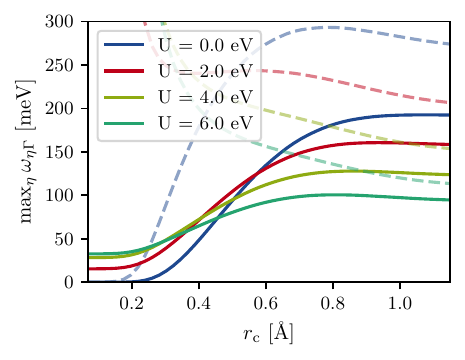}
     \caption{Highest (minority) magnon energy at the $\Gamma$-point calculated as a function of site radius $r_\mathrm{c}$ for different values of U. The solid lines show the magnon energy obtained with a fixed value of $S=5/2$ and the dashed lines show the energies using $S(r_\mathrm{c})$ at the majority sites evaluated from Eq. \eqref{eq:spin_operator_av}. The magnon energies only reach flat plateau (indicating a well-defined site-based model) if we fix $S$ in the calculations.}
     \label{fig:J_r}
\end{figure}
In Fig. \ref{fig:J_r} we show the highest magnon energy at the $\Gamma$-point as a function of the radius of the site $r_\mathrm{c}$ for different values of U. As mentioned above, the magnon energies decrease with increasing values of U. This may be rationalized by noting that Hubbard corrections tend to localize states (improve the delocalization error), which tends to decrease orbital overlaps and thus the exchange interactions. Alternatively, the trend can be understood from the fact that LDA+U increases the band gap, which leads to a reduced response through Eq. \eqref{eq:chi^+-_KS}. In any case it is observed that magnon energies become largely insensitive to the site radius when it exceeds $\sim$ 0.9--1.0 {\AA} and we regard this as formal justification for applying a site-based model to the problem.

In the calculations above, we have mapped the system to a Heisenberg model with $S=5/2$. However, the sites used in the model are spheres centered on the Fe atoms, meaning that the magnetic moment encapsulated within the sites themselves is smaller than the nominal 5 $\mu_\mathrm{B}$ (the sites do not span the entire crystal). 
\begin{figure}[tb]
     \centering
     \includegraphics[scale=1.0]{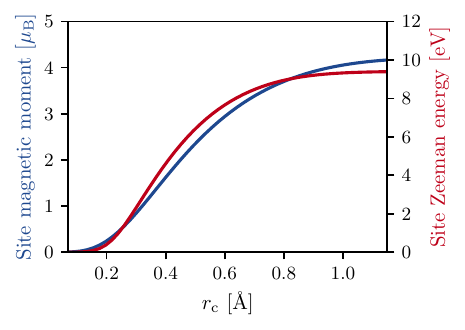}
     \caption{Integrated site magnetization $m^a$ and site Zeeman energy $E_\mathrm{Z}^a$ of the majority site calculated as a function of site radius $r_\mathrm{c}$.}
     \label{fig:site_zeeman}
\end{figure}
In Fig. \ref{fig:site_zeeman} we show the local magnetic moment of the tetrahedrally coordinated Fe atoms (calculated from Eq.  \eqref{eq:spin_operator_av}) as a function of the site radius. Although the magnetization density is approximately localized within $\sim1.0$ {\AA} of the Fe nuclei, the site moments do not reach a flat plateau because a dilute (but significant) fraction of magnetization resides on the ligands and in the interstitial regions. This means that we cannot unambiguously assign a value to $S$ based on the magnetization density alone. The exchange constants \eqref{eq:Liechtenstein}, on the other hand, do not rely explicitly on the magnetization density, but rather on the product of pairs of wave functions with the exchange-correlation magnetic field, which all-in-all exhibits a stricter localization to the sites. This can be illustrated in terms of the site Zeeman energy, 
\begin{equation}\label{eq:site_zeeman}
    E_\mathrm{Z}^a = -\int_{\Omega^a}d\mathbf{r}\, B^\mathrm{xc}(\mathbf{r})m(\mathbf{r}),
\end{equation}
which is shown in Fig. \ref{fig:site_zeeman}. % for the tetrahedrally coordinated Fe sites as a function of the spherical site radius $r_\mathrm{c}$. 
Similar to the total magnon band width in Fig. \ref{fig:J_r}, the site Zeeman energy exhibits a flat plateau for $r_\mathrm{c}>1.0$. In this way, the \textit{energy} contribution derived from the magnetization of the Fe sites is more strongly localized than the magnetization itself. In essence, this is why it is possible to calculate well-defined site-based exchange couplings at all. %This signifies the strongly localized nature of the exchange-correlation magnetic field, which in turn renders the exchange constants insensitive to the site radius.

The oxidation state (Fe$^{3+}$) and Hund's rule point to a spin of $S=5/2$ although the magnetization of the site in Fig. \ref{fig:site_zeeman} implies that the spin of the site is roughly $4/2$. 
The underlying assumption is then that the diluted magnetization residing outside the site will follow the site moment adiabatically in such a way that the sites effectively behave as having $S=5/2$ for all practical purposes. In particular, since the calculated exchange parameters appear to reach a constant value beyond a certain radius, it is highly plausible that the domains $\Omega^a$ could be extended to cover the entire crystal in such a way that each site integrates to $S=5/2$ {\it without} affecting the spin renormalized exchange parameters $\tilde{J}^{ab}_\mathbf{q}$. In contrast, if we choose to define $S$ from Eq. \eqref{eq:spin_operator_av}, the magnon dispersion becomes overly sensitive to the choice of $r_\mathrm{c}$---see Fig. \ref{fig:J_r}---and it will be difficult to regard the site-based model as well defined. %Experimentally, fitting the exchange constants entails a choice of $S^a$ as well, and it is important that the value of $S^a$ is reported and used in conjunction with the associated exchange parameters. For example, the exchange parameters in Ref. \cite{CHEREPANOV199381} was obtained with a choice of $S^a_\mathrm{Cl}=\sqrt(S(S+1))$, which corresponds to local moments of 5.96$\mu_\mathrm{B}$. Such a choice implies that the magnitude of fitted exchange parameters are reduced compared to the choice of $S^a$, but the magnon energies are of course independent of $S^a$ as long as the value is used consistently in both the fitting and calculations (and we used this value for calculating the Cherepanov magnon dispersion in Fig. \ref{fig:YIG_magnons}). However, in our opinion, any choice of $S^a$ that is not half-integer is conceptually wrong for insulators.

\section{Outlook}
We have introduced a new implementation for calculating exchange parameters within the magnetic force theorem. It allows seamless computations directly from the Kohn-Sham orbitals for any system where standard DFT calculations can be performed. As an example, we applied it to the important case of YIG, where the 20 Fe atoms were chosen as magnetic sites. The primitive unit cell contains a total of 80 atoms and it would be very demanding to store the two-point susceptibility $\chi_\mathrm{KS}^{+-}$ in memory using plane waves (with a reasonable cutoff energy). It would also be challenging to generate the Wannier function for the 400 bands required in the calculation, and transforming a plane wave calculation to a localized basis set is therefore rather impractical. In contrast, the present approach only requires evaluation of the integrals \eqref{eq:dxc nks}, which, in principle, can be achieved in any basis set.

Once the site spin pair energies \eqref{eq:dxc nks} have been implemented, the method is largely "plug-and-play". This makes the present approach robust and very suitable for high-throughput calculations of adiabatic magnon spectra. These may in turn be used to predict critical temperatures \cite{pavizhakumari_beyond_2025,rajeev_pavizhakumari_predicting_2025} for magnetic order and strongly facilitate {\it ab initio} screening studies for new room-temperature magnets. Alternatively, the calculated magnon spectrum may be used to determine the magnetic ground state of a material \cite{tellez-mora_systematic_2024} or to extract magnon-phonon coupling parameters \cite{fang_efficient_2025}, which requires calculations of exchange parameters for a multitude of atomic displacements.

\begin{acknowledgments}
The authors acknowledge funding from the Villum foundation Grant No.~00029378. Computations were performed on Nilfheim, a high performance computing cluster at the Technical University of Denmark. We acknowledge support from the Novo Nordisk Foundation Data Science Research Infrastructure 2022 Grant: A high-performance computing infrastructure for data-driven research on sustainable energy materials, Grant no. NNF22OC0078009. The authors wish to thank Mikael Kuisma for fruitful discussions regarding implementation in the early stage of the project.
\end{acknowledgments}

\appendix

\section{GPAW implementation}\label{sec:gpaw implementation}

%By implementing functionality to calculate matrix elements \eqref{eq:site spin pair energy} within the GPAW linear response time-dependent density functional theory code \cite{Yan2011,Skovhus2021,Mortensen2024}, we can straightforwardly use the code's generalized susceptibility framework to calculate the exchange constants \eqref{eq:exchange as linear response} with all-electron accuracy, thanks to the underlying projector augmented wave (PAW) method.
%In order to enable calculations of reciprocal space exchange interactions \eqref{eq:Jt_q} within the GPAW electronic structure code, we have implemented functionality to calculate the site spin pair energies \eqref{eq:dxc nks} in the projector augmented wave (PAW) method. With these in hand, $\tilde J^{ab}_{ij}$ is calculated within GPAW's generalized susceptibility framework \cite{Yan2011,Skovhus2021,Mortensen2024}, summing over all relevant band and $\mathbf{k}\rightarrow\mathbf{k}+\mathbf{q}$ transitions.
Here we provide a detailed account of how the site spin pair energies \eqref{eq:dxc nks} are evaluated within the PAW framework, with application to the GPAW code \cite{Mortensen2024}. We start by introducing the PAW formalism and then discuss how local properties of atomic sites (spherical integrals) can be evaluated accurately using a smooth cutoff function. Finally, we provide the explicit expressions required to calculate \eqref{eq:dxc nks}.

\subsection{Projector augmented wave method}
The projector augmented wave method (PAW) \cite{blochl} provides a formal framework for representing wave functions within a dual basis of smooth pseudowaves spanning all of space and hard all-electron partial waves restricted to augmentation spheres around each atom of the crystal. For a collinear set of Kohn-Sham orbitals $\psi_{ns}$, the all-electron single-particle wave functions are related to smooth pseudo counterparts $\tilde{\psi}_{ns}$ via the linear mapping
\begin{equation}\label{eq:paw mapping}
    |\psi_{ns}\rangle = \hat{\mathcal{T}} |\tilde{\psi}_{ns}\rangle.
\end{equation}
The mapping consists of a set of all-electron partial waves $\phi_i^a$ for each atom $a$ with corresponding smooth partial waves $\tilde{\phi}_i^a$ satisfying $\phi_i^a(\mathbf{r}) - \tilde{\phi}_i^a(\mathbf{r})=0$ for $\mathbf{r}$ outside the augmentation sphere, $r>r_\mathrm{aug}$. Using projector functions $\tilde{p}_i^a$ (satisfying $\sum_i|\tilde{\phi}_i^a\rangle\langle\tilde{p}_i^a| = 1$ within the augmentation sphere) %to project the smooth pseudo waves $\tilde{\psi}_{ns}$ onto the smooth partial waves $\tilde{\phi}_i^a$, 
the smooth part of the wave function is replaced with a hard all-electron counterpart inside each augmentation sphere:
\begin{equation}\label{eq:paw T operator}
    \hat{\mathcal{T}} = 1 + \sum_a \sum_i \left(|\phi_i^a\rangle - |\tilde{\phi}_i^a\rangle\right)\langle\tilde{p}_i^a|.
\end{equation}
In order to formalize this mapping, the partial waves $\phi_i^a$ are taken to be centered at the atom position $\mathbf{R}^a$ such that $\langle\mathbf{r}|\phi_i^a\rangle = \phi_i^a(\mathbf{r}-\mathbf{R}^a)$. 
Assuming that the augmentation spheres do not overlap, one can then use a smooth basis set to represent $\tilde{\psi}_{ns}$ and calculate all-electron matrix elements of arbitrary quasilocal operators $\hat{A}$ directly via a pseudo contribution and a PAW correction:
\begin{equation}\label{eq:paw matrix elements}
    \langle\psi_{ns}|\hat{A}|\psi_{ms'}\rangle = \langle\tilde{\psi}_{ns}|\hat{A}|\tilde{\psi}_{ms'}\rangle + \Delta A_{ns,ms'}.
\end{equation}
The PAW correction is itself calculated from the projector overlaps,
\begin{equation}\label{eq:paw correction}
    \Delta A_{ns,ms'} = \sum_a \sum_{i,i'} \langle\tilde{\psi}_{ns}|\tilde{p}_i^a\rangle \Delta A^a_{ii'}  \langle\tilde{p}^a_{i'}|\tilde{\psi}_{ms'}\rangle,
\end{equation}
and the correction tensor $\Delta A^a_{ii'}$ can be precalculated based on the input basis of all-electron and smooth partial waves,
\begin{equation}
    \Delta A^a_{ii'} \equiv \langle \phi_i^a| \hat{A} |\phi^a_{i'}\rangle - \langle \tilde{\phi}_i^a| \hat{A} |\tilde{\phi}^a_{i'}\rangle.
\end{equation}

\subsection{Local properties of atomic sites}

Before tackling the site spin pair energies \eqref{eq:dxc nks}, it is worth taking a step back and consider how to calculate site properties which can be written as simpler functionals of the electron density $n$ and magnetization $\mathbf{m}$ within the PAW method. For collinear magnetic systems, it is of particular relevance to calculate the site magnetization,
\begin{equation}\label{eq:site magnetization}
    n_z^a = \int d\mathbf{r}\, \Theta(\mathbf{r} \in \Omega^a) n_z(\mathbf{r})
\end{equation}
with $n_z=n_\uparrow-n_\downarrow$, and the site Zeeman energy,
\begin{equation}\label{eq:site zeeman energy}
    E_\mathrm{Z}^a = - \int d\mathbf{r}\, \Theta(\mathbf{r} \in \Omega^a) W_z^\mathrm{xc}(\mathbf{r}) n_z(\mathbf{r})
\end{equation}
with $W_z^\mathrm{xc}=(V_\uparrow^\mathrm{xc} - V_\downarrow^\mathrm{xc})/2$, where $\Omega^a$ denotes the site volume of sublattice $a$ centered in the first unit cell. As described in Sec. \ref{sec:theory}, the site magnetization can be used to define the length of the spin operators \eqref{eq:spin_operator} in the Heisenberg model \eqref{eq:heisenbrghamiltonian}, see Eq. \eqref{eq:spin_operator_av}, and the site Zeeman energy quantifies the local exchange splitting of sublattice $a$ in terms of how much each site lowers the total energy due to the exchange-correlation Zeeman term. 

Now, given a specific approximation to the exchange-correlation functional, both Eqs. \eqref{eq:site magnetization} and \eqref{eq:site zeeman energy} can be rewritten in terms of a known density functional $f[n,\mathbf{m}]$,
\begin{equation}
    f^a = \int d\mathbf{r}\, \Theta(\mathbf{r} \in \Omega^a) f[n,\mathbf{m}](\mathbf{r}).
\end{equation}
If the site volume $\Omega^a$ is chosen such that it overlaps only with the augmentation sphere of atom $a$ and the functional $f$ is semilocal, evaluation of such a site property can be split into a pseudo contribution
\begin{equation}\label{eq:site property pseudo contribution}
    \tilde{f}^a = \int d\mathbf{r}\, \Theta(\mathbf{r} \in \Omega^a) f[\tilde{n},\tilde{\mathbf{m}}](\mathbf{r})
\end{equation}
and a single-atom PAW correction,
\begin{equation}\label{eq:site property PAW correction}
    \Delta f^a = \int_{\Omega^a_\mathrm{aug}} d\mathbf{r}\, \Theta(\mathbf{r} \in \Omega^a) \left(f[n,\mathbf{m}](\mathbf{r}) - f[\tilde{n},\tilde{\mathbf{m}}](\mathbf{r})\right),
\end{equation}
where the pseudo density $\tilde{n}$ and pseudo magnetization $\tilde{\mathbf{m}}$ are defined as the pseudo contributions to the density $n$ and magnetization $\mathbf{m}$ respectively, see Eq. \eqref{eq:paw matrix elements}. The %it is straightforward to evaluate augmentation sphere integrals such as Eq. \eqref{eq:site property PAW correction}. 
evaluation of \eqref{eq:site property PAW correction} is straightforward. In GPAW, each augmentation sphere is represented by a dense nonlinear radial grid and a Lebedev quadrature of degree 11 meaning that angular functions of polynomial order 11 are integrated exactly. However, in order to integrate pseudo contributions \eqref{eq:site property pseudo contribution} one needs to use an equidistant real-space grid, which is related to a corresponding plane-wave representation via a fast Fourier transform. Using this grid, smooth quantities such as $\tilde{n}$ are integrated accurately, but even if it is safe to assume that $f[\tilde{n},\tilde{\mathbf{m}}](\mathbf{r})$ is sufficiently smooth, the Heavyside step function $\Theta$ is not.

In order to define the magnetic sites \eqref{eq:spin_operator} in a way which permits efficient evaluation of pseudo contributions \eqref{eq:site property pseudo contribution} on an equidistant real-space grid, we therefore redefine $\Theta(\mathbf{r} \in \Omega^a)$ as a smooth step-like truncation function with spherical symmetry. Centered at the atom position $\mathbf{R}^a$, the radial truncation function $\Theta(r)$ is chosen to smoothly interpolate between values
\begin{equation}\label{eq:radial truncation function constraints}
    \Theta(r) = 
    \begin{cases}
        1 & \text{for $r \leq r_\mathrm{c} - \Delta r_\mathrm{c}/2$},\\
        \lambda & \text{if $r=r_\mathrm{c}$},\\
        0 & \text{for $r \geq r_\mathrm{c} + \Delta r_\mathrm{c}/2$},
    \end{cases}
\end{equation}
with $r_\mathrm{c}$ and $\Delta r_\mathrm{c}$ representing the spherical cutoff radius and cutoff interpolation range respectively, while $0<\lambda<1$. The smooth transition from unity to zero is accomplished by the function
\begin{align}\label{eq:radial truncation function}
    \Theta(r) = 
    %&\frac{\tau\left(\frac{1}{2} - \frac{r - r_\mathrm{c}}{\Delta r_\mathrm{c}}\right)}{\tau\left(\frac{1}{2} - \frac{r - r_\mathrm{c}}{\Delta r_\mathrm{c}}\right) + \left(\frac{1}{\lambda} - 1\right) \tau\left(\frac{1}{2} + \frac{r - r_\mathrm{c}}{\Delta r_\mathrm{c}}\right)} \\
    &\left[1 + \left(\frac{1}{\lambda}-1\right) \frac{\tau\left(\frac{1}{2} + \frac{r - r_\mathrm{c}}{\Delta r_\mathrm{c}}\right)}{\tau\left(\frac{1}{2} - \frac{r - r_\mathrm{c}}{\Delta r_\mathrm{c}}\right)} \right]^{-1} \\
    &\hspace{70pt} \text{for $r_\mathrm{c} - \Delta r_\mathrm{c}/2 < r < r_\mathrm{c} + \Delta r_\mathrm{c}/2$}, \nonumber
\end{align}
where
\begin{equation}
    \tau(x) =
    \begin{cases}
        0 & \text{for $x<=0$},\\
        e^{-1/x} & \text{for $x>0$}.
    \end{cases}
\end{equation}
By construction, $\Theta(r)$ and its derivatives are continuous, 
\begin{figure}[tb]
    \centering
    \includegraphics[scale=1]{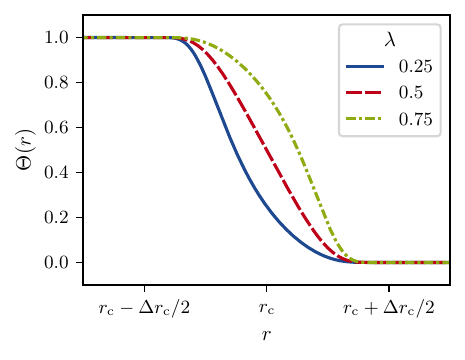}
    \caption{Radial truncation function $\Theta(r)$ as defined by Eqs. \eqref{eq:radial truncation function constraints} and \eqref{eq:radial truncation function} for different values of $\lambda$.}
    \label{fig:spherical_truncation_function}
\end{figure}
providing a well-controlled step-like truncation as illustrated in Fig. \ref{fig:spherical_truncation_function}. In order to strengthen the interpretation of $r_\mathrm{c}$ as a spherical cutoff radius, we fix $\lambda$ to achieve an effective site volume of
\begin{equation}\label{eq:atomic site volume}
    4\pi \int_0^\infty r^2 dr\, \Theta(r) = \frac{4\pi}{3} r_\mathrm{c}^3.
\end{equation}
Furthermore, we fix the cutoff interpolation range $\Delta r_\mathrm{c}$ based on the grid point volume $\delta V$ associated with the coarse equidistant real-space grid, $\Delta r_\mathrm{c} = \delta V^{1/3}$. With this choice, we are able to recover the site volume \eqref{eq:atomic site volume} with less than 1\% error when integrating $\Theta(\mathbf{r} \in \Omega^a)$ on the equidistant real-space grid for a wide range of cutoffs $r_\mathrm{c}$ (setting $f=1$ in Eq. \eqref{eq:site property pseudo contribution}). The cutoff radius $r_\mathrm{c}$ is left as the central parameter that defines the magnetic sites such that local site properties such as magnetic moments or site spin Zeeman energies can be computed as {\it functions} of $r_\mathrm{c}$. 

\subsection{Site matrix elements}

For the site spin pair energy \eqref{eq:dxc nks}, the pseudo contribution \eqref{eq:paw matrix elements} is given by
\begin{equation}\label{eq:dxc pseudo contribution}
    \tilde{d}^{\mathrm{xc},a}_{n\mathbf{k}s,m\mathbf{k}'s'} = \int d\mathbf{r}\, \tilde{\psi}_{n\mathbf{k}s}^*(\mathbf{r}) \Theta(\mathbf{r} \in \Omega^a) B^\mathrm{xc}(\mathbf{r}) \tilde{\psi}_{m\mathbf{k}'s'}(\mathbf{r}).
\end{equation}
Assuming $B^\mathrm{xc}(\mathbf{r})$ to be sufficiently smooth, $\tilde{d}^{\mathrm{xc},a}_{n\mathbf{k}s,m\mathbf{k}'s'}$ can be accurately integrated on the coarse equidistant real-space grid, thanks to the smooth definition \eqref{eq:radial truncation function} of the spherical atomic sites. What remains is the PAW correction \eqref{eq:paw correction}, which is calculated from the projector overlaps $\langle\tilde{p}^a_i|\tilde{\psi}_{n\mathbf{k}s}\rangle$ and the PAW correction tensor,
\begin{align}\label{eq:dxc paw correction tensor def}
    \Delta d^{\mathrm{xc},a}_{ii'} = &\langle \phi_i^a| \Theta(\mathbf{r} \in \Omega^a)\, B^{\mathrm{xc}}(\mathbf{r}) |\phi^a_{i'}\rangle \nonumber\\
    &- \langle \tilde{\phi}_i^a| \Theta(\mathbf{r} \in \Omega^a)\, B^{\mathrm{xc}}(\mathbf{r}) |\tilde{\phi}^a_{i'}\rangle.
\end{align}
The partial waves are represented by real radial functions $\phi_i^a(r)$ defined on the nonlinear radial grid of the augmentation sphere and a real spherical harmonic,
\begin{equation}
    \phi_i^a(\mathbf{r}) = Y_{l_i}^{m_i}(\hat{\mathbf{r}}) \phi_i^a(r), \quad \tilde{\phi}_i^a(\mathbf{r}) = Y_{l_i}^{m_i}(\hat{\mathbf{r}}) \tilde{\phi}_i^a(r).
\end{equation}
In the vicinity of atom $a$, also $B^\mathrm{xc}(\mathbf{r})$ can be expanded in terms of spherical harmonics (centered on the atom):
\begin{equation}
    B^{\mathrm{xc},a}(\mathbf{r}) \simeq \sum_{l=0}^{l_\mathrm{max}} \sum_{m=-l}^l Y_l^m(\hat{\mathbf{r}}) B^{\mathrm{xc},a}_{lm}(r).
\end{equation}
We choose $l_\mathrm{max}=4$, for which the expansion coefficients $B^{\mathrm{xc},a}_{lm}(r)$ can be evaluated exactly on the Lebedev quadrature of degree 11. % of our chosen $l_\mathrm{max}=4$. 
Utilizing the spherical symmetry of the truncation function $\Theta(r)$, % which is centered at the atom position $\mathbf{R}^a$ and assumed to overlap only with augmentation sphere $a$, 
the PAW correction tensor \eqref{eq:dxc paw correction tensor def} can then be calculated as
\begin{align}\label{eq:dxc paw correction tensor}
    \Delta d^{\mathrm{xc},a}_{ii'} 
    = \sum_{l=0}^{l_\mathrm{max}} \sum_{m=-l}^l &g_{l,l_i,l_{i'}}^{m,m_i,m_{i'}} \int r^2 dr\, \Theta^a(r) B^{\mathrm{xc},a}_{lm}(r) \nonumber \\
    &\times \left[\phi_i^a(r) \phi_{i'}^a(r) - \tilde{\phi}_i^a(r) \tilde{\phi}_{i'}^a(r)\right],
\end{align}
where the radial integral is carried out on the nonlinear grid of the augmentation sphere,  $\Theta^a(r)=\Theta(|\mathbf{r}-\mathbf{R}^a|)$ and the Gaunt coefficients $g_{l,l_i,l_{i'}}^{m,m_i,m_{i'}}$ yield the integral over three spherical harmonics. 

This completes our implementation for calculating exchange constants \eqref{eq:Jt_q} in reciprocal space. The main assumption is that $B^\mathrm{xc}(\mathbf{r})$ is sufficiently smooth to permit evaluation of the pseudo contribution \eqref{eq:dxc pseudo contribution} on a coarse equidistant real-space grid. This may seem a crude approximation, but the validity of the resulting matrix elements \eqref{eq:dxc nks} can be affirmed by comparing the site Zeeman energy \eqref{eq:site zeeman energy} calculated from Eqs. \eqref{eq:site property pseudo contribution} and \eqref{eq:site property PAW correction} to the sum rule
\begin{align}
    E^{ab}_\mathrm{Z}(\mathbf{q}) = &\frac{1}{N_k} \sum_{\mathbf{k}}\sum_{n,m} \left(f_{n\mathbf{k}\uparrow} - f_{m\mathbf{k}+\mathbf{q}\downarrow}\right) \nonumber \\
    &\times E^{a}_{\mathrm{Z},n\mathbf{k}\uparrow,m\mathbf{k}+\mathbf{q}\downarrow}n^{b}_{m\mathbf{k}+\mathbf{q}\downarrow,n\mathbf{k}\uparrow}
    = \delta^{ab} E_\mathrm{Z}^a,
\end{align}
where the site pair Zeeman energy and site pair density, given by
\begin{subequations}
    \begin{equation}
        E^{a}_{\mathrm{Z},n\mathbf{k}s,m\mathbf{k}'s'} = 
    -\langle \psi_{n\mathbf{k}s}| \Theta(\mathbf{r} \in \Omega^a)\, W^{\mathrm{xc}}(\mathbf{r}) |\psi_{m\mathbf{k}'s'}\rangle
    \end{equation}
    and
    \begin{equation}
        n^{a}_{n\mathbf{k}s,m\mathbf{k}'s'} = 
    \langle \psi_{n\mathbf{k}s}| \Theta(\mathbf{r} \in \Omega^a) |\psi_{m\mathbf{k}'s'}\rangle
    \end{equation}
\end{subequations}
respectively, are calculated analogously to Eqs. \eqref{eq:dxc pseudo contribution} and \eqref{eq:dxc paw correction tensor}. Provided that the spherical cutoff radius $r_\mathrm{c}$ properly enclose the $d$-orbitals of the magnetic atoms of a given crystal, we find this sum rule to be fulfilled, which largely validates the implementation.% It  is appropriate as well as flexible ($r_\mathrm{c}$ can be varied independently for each atom) and straightforward (Eq. \eqref{eq:Jt_q} can be evaluated directly from a DFT ground state calculation without any additional mapping to a tight-binding model etc.).

\section{Including Hubbard U in MFT calculations}\label{sec:mft_U}
Hubbard corrections are most naturally incorporated in the  MFT formula for exchange constants using an orbital-based definition of the sites. For sites defined in real space the inclusion is more tricky because the Hubbard corrections give rise to a nonlocal contribution to the exchange-correlation potential. At the end of the day, however, the Hubbard potential results only in corrections to the site spin pair energies which are straightforward to approximate. 

%In this appendix, we start by reviewing how Hubbard corrections effectively introduce nonlocal potentials to DFT. Subsequently, we derive a formula for the DFT+U exchange constants in the continuum MFT approach and show how these can be implemented in the PAW method.

\subsection{Hubbard corrections in the Dudarev form}
For a general noncollinear system, the Hubbard correction in the Dudarov approach is given by \cite{PhysRevB.57.1505}:
\begin{align}
E_U=\frac{1}{2}\sum_a U^a \sum_{i,j}\sum_{\sigma,\sigma'}(\rho_{ij}^{a,\sigma\sigma'}\delta_{ij}\delta^{\sigma\sigma'}-\rho_{ij}^{a,\sigma\sigma'}\rho_{ji}^{a,\sigma'\sigma}).
\end{align}
Here
\begin{align}\label{eq:occupation matrix}
\rho_{ij}^{a,\sigma\sigma'}=\sum_nf_n\langle\phi_{i}^a|\psi_n^{\sigma}\rangle\langle\psi_n^{\sigma'}|\phi_{j}^a\rangle
\end{align}
denotes the occupation matrix for the atomic orbitals $\phi^a_i$ of atom $a$, $i$ and $j$ iterate the atom-specific orbital subspace subject to Hubbard corrections (typically valence $d$ or $f$ orbitals) and $\sigma,\sigma'$ are spinor indices. %atom $a$, $\sigma$ and $\sigma'$ are spinor indices and $\phi^a_i$ is an atomic orbital $i$ belonging to atom $a$. The sums over $i,j$ only run over the atomic orbitals that are to receive Hubbard corrections. 
%We note that this can be written as
%\begin{align}
%\rho_{a,ij}^{\sigma\sigma'}=\int d\mathbf{r}d\mathbf{r}'\phi^*_{a,i}(\mathbf{r})\rho^{\sigma\sigma'}(\mathbf{r}, \mathbf{r}')\phi_{a,j}(\mathbf{r}')
%\end{align}
%with
%\begin{align}
%\rho^{\sigma\sigma'}(\mathbf{r},\mathbf{r}')=\sum_nf_n\psi_n^{\sigma}(\mathbf{r})\psi_n^{\sigma'*}(\mathbf{r}').
%\end{align}
The corresponding (nonlocal) contribution to the Kohn-Sham potential is
\begin{align}
V^{a,\sigma\sigma'}_{ij}&=\frac{\partial E_U}{\partial \rho_{ji}^{a,\sigma'\sigma}}\notag\\&=\frac{U^a}{2}(\delta_{ij}\delta^{\sigma\sigma'}-2\rho_{ij}^{a,\sigma\sigma'})\notag\\
&=\frac{U^a}{2}[(\delta_{ij}-n^a_{ij})\delta^{\sigma\sigma'}-\mathbf{m}_{ij}^a\cdot\boldsymbol{\sigma}^{\sigma\sigma'}],
\end{align}
where the occupation matrix has been rewritten in terms of the atomic orbital density $n^a_{ij}=\mathrm{Tr}[\rho^a_{ij}]$ and magnetization $\mathbf{m}_{ij}^a=\mathrm{Tr}[\boldsymbol{\sigma}\rho^a_{ij}]$. %It may also be written in real space as
The Kohn-Sham potential is nonlocal and operates on a spinor wavefunction $\psi(\mathbf{r})$ as $\int d\mathbf{r}'\, V^a(\mathbf{r}, \mathbf{r}') \psi(\mathbf{r}')$, where the real-space representation of the potential is given by:
\begin{align}
V^a(\mathbf{r},\mathbf{r}')=\frac{U^a}{2}\sum_{i,j}\phi^a_{i}(\mathbf{r})\phi_{j}^{a*}(\mathbf{r}')[(\delta_{ij}-n^a_{ij})-\mathbf{m}_{ij}^a\cdot\boldsymbol{\sigma}].
\end{align}
%which emphasizes the non-local nature of the potential. 
The magnetic part of the potential can thus be regarded as a functional of the nonlocal atomic magnetization densities, $\mathbf{m}^a(\mathbf{r},\mathbf{r}')=\sum_{ij}\phi^a_{i}(\mathbf{r})\mathbf{m}^a_{ij}\phi^{a*}_{j}(\mathbf{r}')$. 

\subsection{Exchange constants in DFT+U}

Assuming that the direction of magnetization is constant within each atomic site (rigid spin approximation), we may consider the effect of a %rotation that depends on the site, but are independent of $\mathbf{r}$ and $\mathbf{r}'$. We then consider a 
small sitewise rotation of the magnetization direction $\delta \mathbf{u}^a$. Such a rotation yields a perturbing potential
\begin{align}
\delta V^{a}(\mathbf{r},\mathbf{r}')=\Big[
&\Theta(\mathbf{r} \in \Omega^a) B^\mathrm{xc}(\mathbf{r})\delta(\mathbf{r}-\mathbf{r}')
\nonumber \\
&-\frac{U^a}{2}m^a(\mathbf{r},\mathbf{r}')\Big]\delta\mathbf{u}^a\cdot\boldsymbol{\sigma},
\end{align}
where $m^a(\mathbf{r},\mathbf{r}')$ is the magnitude of $\mathbf{m}^a(\mathbf{r},\mathbf{r}')$, and the Hubbard corrected orbital subspace is assumed to be effectively enclosed inside of the site volume $\Omega^a$. The transverse contribution to the energy difference induced by this rotation is then given by the magnetic force theorem \cite{Durhuus2023} as
\begin{align}\label{eq:transverse dE}
\Delta E^{(2, \perp)}=\sum_{a,b}\sum_n\sum_{m\neq n}f_n\frac{\alpha_{nm}^a\alpha_{mn}^{b}}{\varepsilon_n -\varepsilon_m},
\end{align}
where
\begin{align}\label{eq:alpha}
\alpha_{nm}^a&=\langle\psi_n|\Big(\hat{\Theta}^a\hat B^\mathrm{xc}
-\frac{U^a}{2}
%\sum_{i,j}|\phi_{i}^a\rangle m_{ij}^a\langle\phi_{j}^a|
\hat{m}^a
\Big)\boldsymbol{\sigma}|\psi_m\rangle\cdot\delta\mathbf{u}_\perp^{a},
%\notag\\
%&=\Big(\langle\psi_m|\hat B^\mathrm{xc}\boldsymbol{\sigma}|\psi_n\rangle-\frac{U}{2}\sum_{i,j}m_{ij}^aP_{mi}^{a\dag}\boldsymbol{\sigma}P_{nj}^a\Big)\cdot\delta\mathbf{u}_\perp^{a}.
\end{align}
%Here $P_{ni}^a=\langle\phi^a_i|\psi_n\rangle$ denote the spinor projections of the Kohn-Sham states onto the atomic orbitals 
%$m_{ij}^a$ are the expansion coefficients of $m^a(\mathbf{r},\mathbf{r}')$ in a site-localized orbital basis 
and $\delta\mathbf{u}_\perp^{a}$ is the transverse component of the change in magnetization direction (linear in the rotation angle $\theta$).

%In the absence of spin-orbit coupling, we can take the eigenstates to be collinearly 
For a collinear system (absent of spin-orbit coupling), we can take $|\psi_n\rangle \rightarrow |\psi_{ns}\rangle$ with $s\in\{\uparrow,\downarrow\}$, and consider a transverse change in magnetization direction $\delta\mathbf{u}_\perp^{a}=\theta^a (\cos\phi^a, \sin\phi^a, 0)$. Insertion into the transverse energy difference \eqref{eq:transverse dE} then yields
\begin{align}\label{eq:transverse dE nosoc}
\Delta E^{(2, \perp)} = \sum_{a,b}\sum_{n,m}
    &\frac{f_{n\uparrow} - f_{m\downarrow}}
    {\varepsilon_{n\uparrow} -\varepsilon_{m\downarrow}} 
    d_{n\uparrow,m\downarrow}^{\mathrm{xc},a} d_{m\downarrow,n\uparrow}^{{\mathrm{xc},b}}
    \nonumber \\
    &\times \theta^a \theta^b \left(\cos\phi^a \cos\phi^b + \sin\phi^a\sin\phi^b\right),
\end{align}
where
\begin{align}\label{eq:hubbard corrected dxc}
d_{ns,ms'}^{\mathrm{xc},a}&=\langle\psi_{ns}|\Big(\hat{\Theta}^a\hat B^\mathrm{xc}
-\frac{U^a}{2}
%\sum_{i,j}|\phi_{i}^a\rangle m_{ij}^a\langle\phi_{j}^a|
\hat{m}^a
\Big)|\psi_{ms'}\rangle.
\end{align}
Comparing this to the energy difference arising from the transverse components in the Heisenberg model, see Eq. \eqref{eq:heisenbrghamiltonian_av}, it is thus clear that the MFT equation \eqref{eq:exchange as linear response} still holds, but now with a Hubbard correction to the DFT site spin pair energy,
\begin{equation}
d^{\mathrm{xc},a}_{ns,ms'}\rightarrow d^{\mathrm{xc},a}_{ns,ms'}+\Delta d^{\mathrm{xc},a}_{ns,ms'},
\end{equation}
where $\Delta d^{\mathrm{xc},a}_{ns,ms'}$ is proportional to $U^a$. In order to calculate this correction one needs, in principle, to take the absolute value $|\mathbf{m}^a(\mathbf{r},\mathbf{r}')|$ in real space. This is rather impractical, since 
it entails the need to evaluate nonlocal matrix elements $\iint d\mathbf{r}d\mathbf{r}'\: \psi_{ns}^*(\mathbf{r}) m^a(\mathbf{r},\mathbf{r}') \psi_{ms'}(\mathbf{r}')$ using some real-space representation for $m^a$.
%evaluate matrix elements $ \langle\psi_{ns}|\hat{O}|\psi_{ms'}\rangle$ for \textit{arbitrary} nonlocal operators $\hat{O}$. 
However, in the spirit of the rigid spin approximation, $\mathbf{m}^a(\mathbf{r},\mathbf{r}') \simeq m^a(\mathbf{r},\mathbf{r}') \mathbf{u}^a$, we may approximate $m^a(\mathbf{r},\mathbf{r}') \simeq \mathbf{m}^a(\mathbf{r},\mathbf{r}') \cdot \mathbf{u}^a$
such that
\begin{equation}\label{eq:dxc hubbard correction}
    \Delta d^{\mathrm{xc},a}_{ns,ms'} \simeq -\frac{U^a}{2}\sum_{i,j} m_{ij}^a \langle \psi_{ns}|\phi_{i}^a\rangle\langle\phi_{j}^a|\psi_{ms'}\rangle,
\end{equation}
where $m_{ij}^a \equiv \mathbf{m}_{ij}^a \cdot \mathbf{u}^a$.

\subsection{Hubbard corrected exchange constants in the PAW method}

%In order to evaluate the Hubbard corrections \eqref{eq:dxc hubbard correction} withi
In the PAW method, we take the Hubbard corrected subspace to consist of all-electron partial waves $\phi_i^a(\mathbf{r})$. Provided that the selected partial waves are well localized within the augmentation sphere of atom $a$, we may then approximate $\langle\phi_i^a|\psi_{ns}\rangle\simeq\langle\tilde{p}_i^a|\tilde{\psi}_{ns}\rangle$. Doing so for the occupation matrix \eqref{eq:occupation matrix} and the overlaps of the correction \eqref{eq:dxc hubbard correction} itself, it becomes straightforward to calculate $\Delta d^{\mathrm{xc},a}_{ns,ms'}$ with
\begin{equation}
    m_{ij}^a = \left(\rho_{ij}^{a,\uparrow\uparrow} - \rho_{ij}^{a,\downarrow\downarrow}\right) \mathbf{e}_z \cdot \mathbf{u}^a.
\end{equation}
%assume that interstitial contributions (in the augmentation sphere sense) to overlaps $\langle\phi_i^a|\psi_{ns}\rangle$ are negligible. Using that $\sum_i|\tilde{\phi}_i^a\rangle\langle\tilde{p}_i^a| = 1$ within the augmentation sphere, application of Eqs. \eqref{eq:paw mapping} and \eqref{eq:paw T operator} then yields the correction
% \begin{equation}
%     \Delta d^{\mathrm{xc},a}_{ns,ms'} = - \frac{U^a}{2} \sum_{i,j} m_{ij}^a \langle\tilde{\psi}_{ns}|\tilde{p}_i^a\rangle \langle \tilde{p}_j^a| \tilde{\psi}_{ms'}\rangle.
% \end{equation}
%For collinear systems, we assume that 
%$m^a(\mathbf{r},\mathbf{r}')=\pm m^a_z(\mathbf{r},\mathbf{r}')$ such that $m_{ij}^a=|n_{ij}^{a,\uparrow} - n_{ij}^{a,\downarrow}|$ with
% \begin{align}
%     n_{ij}^{a,s} =
%     \sum_{n}f_{ns}\langle\tilde{p}_{i}^a|\tilde{\psi}_{ns}\rangle\langle\tilde{\psi}_{ns}|\tilde{p}_{j}^a\rangle.
% \end{align}
Since we are assuming that the Hubbard corrected all-electron partial waves $\phi_i^a(\mathbf{r})$ are well localized inside the augmentation sphere $\Omega_\mathrm{aug}$ as well as 
the chosen site volume $\Omega^a$, one should be a little careful as to what systems and orbitals the correction is applied to. For example, the apparent finite U divergence of the largest magnon energy using $S(r_\mathrm{c})$ in the $r_\mathrm{c}\rightarrow 0$ limit, see Fig. \ref{fig:J_r}, is due to the fact that the Hubbard correction term is held constant regardless of $r_\mathrm{c}$. For this reason, DFT and DFT+U magnon dispersions can only really be compared at the same footing when the site cutoff radius is comparable to or larger than the augmentation sphere radius, which for the default Fe PAW setup in GPAW is two Bohr radii, $r_\mathrm{aug}=2 a_0 \simeq 1.06$ Å.

% The quantity $m^a(\mathbf{r},\mathbf{r}')=|\mathbf{m}^a(\mathbf{r},\mathbf{r}')|$ and thus $m_{ij}^a$ is awkward to deal with. However, assuming a collinear system where the magnetization density is all positive within a site (taken along the $z$-direction) we may take $m_{ij}^a=m_{ij}^{a,z}$. Moreover, in the PAW formalism the magnetization within the augmentation spheres can be expressed as
% \begin{align}
% m_{ij}^{a,z}&=\sum_{n_\uparrow}f_{n_\uparrow}\langle\phi_{i}^a|\psi_{n\uparrow}\rangle\langle\psi_{n\uparrow}|\phi_{j}^a\rangle-\sum_{n_\downarrow}f_{n_\downarrow}\langle\phi_{i}^a|\psi_{n\downarrow}\rangle\langle\psi_{n\downarrow}|\phi_{j}^a\rangle\notag\\
% &\approx\sum_{n_\uparrow}f_{n_\uparrow}p_{n_\uparrow i}^ap^{a*}_{n_\uparrow j}-\sum_{n_\downarrow}f_{n_\downarrow}p^a_{n_\downarrow i}p^{a*}_{n_\downarrow j}.
% \end{align}
% In the second line it was used that $\langle\phi_{i}^a|\psi_{ns}\rangle\approx\langle\tilde p_{i}^a|\tilde\psi_{ns}\rangle\equiv p^a_{nsi}$, where $|\tilde p_{i}^a\rangle$ denotes the projector function associated with orbital $i$ on atom $a$. Finally, if take  $P_{ni}^a\approx p^a_{nsi}$ in Eq. \eqref{eq:alpha} we find that the exchange constants are still given by the expression \eqref{eq:exchange as linear response} and that the site spin pair energies receive the corrections
% \begin{equation}
% d^{\mathrm{xc},ai}_{ns,ms'}\rightarrow d^{\mathrm{xc},ai}_{ns,ms'}+\Delta d^{\mathrm{xc},ai}_{ns,ms'}
% \end{equation}
% where 
% \begin{align}
% \Delta d^{\mathrm{xc},a}_{ns,ms'}=-\frac{U}{2}\sum_{ij}m_{ij}^{a,z}p_{ns i}^{a*}p^a_{ms'j}.
% \end{align}

\bibliographystyle{unsrt}
\bibliography{references}

\end{document}